\documentclass{aa}  

\usepackage{CJK}
\usepackage{graphicx}
\usepackage{float}

\usepackage{color}
\usepackage{txfonts}

\begin{document} 

\begin{CJK*}{UTF8}{gbsn}
   \title{Fitting the radial acceleration relation to individual SPARC galaxies}

   \author{Pengfei Li %（李鹏飞）
       \inst{1}
          \and
          Federico Lelli\inst{2}\fnmsep\thanks{ESO Fellow}
          \and
          Stacy S. McGaugh\inst{1}
          \and
          James M. Schombert\inst{3}
          }

   \institute{Deparment of Astronomy, Case Western Reserve University,
              Cleveland, OH 44106, USA\\
              \email{pengfeili0606@gmail.com}
         \and
             European Southern Observatory, Karl-Schwarschild-Strasse 2, Garching bei M\"{u}nchen, Germany
         \and
         Department of Physics, University of Oregon, Eugene, OR 97403, USA
             }

  \abstract 
  {Galaxies follow a tight radial acceleration relation (RAR): the acceleration observed at every radius correlates with that expected from the distribution of baryons. We use the Markov chain Monte Carlo method to fit the mean RAR to 175 individual galaxies in the SPARC database, marginalizing over stellar mass-to-light ratio ($\Upsilon_{\star}$), galaxy distance, and disk inclination. Acceptable fits with astrophysically reasonable parameters are found for the vast majority of galaxies. The residuals around these fits have an rms scatter of only 0.057 dex ($\sim$13$\%$). This is in agreement with the predictions of modified Newtonian dynamics (MOND). We further consider a generalized version of the RAR that, unlike MOND, permits galaxy-to-galaxy variation in the critical acceleration scale. The fits are not improved with this additional freedom: there is no credible indication of variation in the critical acceleration scale. The data are consistent with the action of a single effective force law. The apparent universality of the acceleration scale and the small residual scatter are key to understanding galaxies.}


   \maketitle
  
%-------------------------------------------------------------------

\section{Introduction} \label{sec:intro}

Since the discovery of the flat rotation curves of disk galaxies \citep{Bosma1978, Rubin1978}, the mass discrepancy problem has been widely explored. The baryonic Tully-Fisher relation \citep[BTFR][]{1977A&A....54..661T, McGaugh2000, Lelli2016} was established as the link between the flat rotation velocity $V_\mathrm{f}$ and the baryonic mass for late-type galaxies. The definition of mass discrepancy at each radius, $\mathrm{M_{tot}/M_{bar} \simeq V^2_{obs}/V^2_{bar}}$, makes it possible to study the ``local'' relation between the rotation curve shape and the baryonic mass distribution, which lead to the mass discrepancy-acceleration relation \citep{McGaugh2004}.

In order to explore the mass discrepancy-acceleration relation further, \citet{2016AJ....152..157L} built the $Spitzer$ Phtometry and Accurate Rotation Curves (SPARC) database: a sample of 175 disk galaxies with homogeneous [3.6] surface photometry and high-quality \rm{H}{\small\rm{I}}/\rm{H}$\alpha$ rotation curves, spanning a wide range in morphological types (S0 to Irr), stellar masses (5 dex), surface brightnesses (4 dex), and gas fractions. Using the SPARC database, \citet{2016PhRvL.117t1101M} established the radial acceleration relation (RAR), in which the observed acceleration ($\mathrm{g_{obs} = V^2_{obs}/R}$) tightly correlates with the baryonic one ($\mathrm{g_{bar}}$). The $\mathrm{g_{obs}}$-$\mathrm{g_{bar}}$ plane has a major advantage over the mass discrepancy-acceleration relation: the two quantities and the corresponding errors are fully independent, thus observed and expected scatters can be easily computed without additional complications from covariances between the measurements. Furthermore, \citet{2017ApJ...836..152L} extend the galaxy sample to include 25 early-type galaxies and 62 dwarf spheroidals, finding that they follow the same relation as late-type galaxies within the uncertainties.  

Assuming that the stellar mass-to-light ratio $\Upsilon_\star$ does not vary strongly at [3.6] \citep{McGaugh2014b, McGaugh2015, Meidt2014, Schombert2014a}, it is found that the RAR has an observed rms scatter of only 0.13 dex \citep{2016PhRvL.117t1101M, 2017ApJ...836..152L}. This is largely driven by uncertainties on galaxy distance and disk inclination, as well as possible galaxy-to-galaxy variations in the value of $\Upsilon_\star$. Hence, the intrinsic scatter around the RAR must be even smaller.

Given that late-type galaxies statistically satisfy the RAR, we can explore its intrinsic scatter by fitting individual galaxies and marginalizing over $\Upsilon_\star$, galaxy distance ($D$), and disk inclination ($i$). This is equivalent to rotation curve fits in modified Newtonian dynamics \citep[MOND,][]{1983ApJ...270..371M}, but our aim here is to measure the intrinsic scatter around the RAR from a purely empirical perspective. Moreover, differently from classic MOND studies \citep[e.g.,][]{Sanders2002}, we impose priors on $\Upsilon_\star$, $D$, and $i$ based on the observational uncertainties. These ``free'' parameters are treated as global quantities for each galaxy, whereas the RAR involves local quantities measured at each radius. Hence, there is no guarantee that adjusting those parameters within the errors can result in satisfactory individual fits for each and every galaxy or decrease the empirical scatter around the mean relation.

\begin{figure*}[t]
\centering
\includegraphics[scale=0.52]{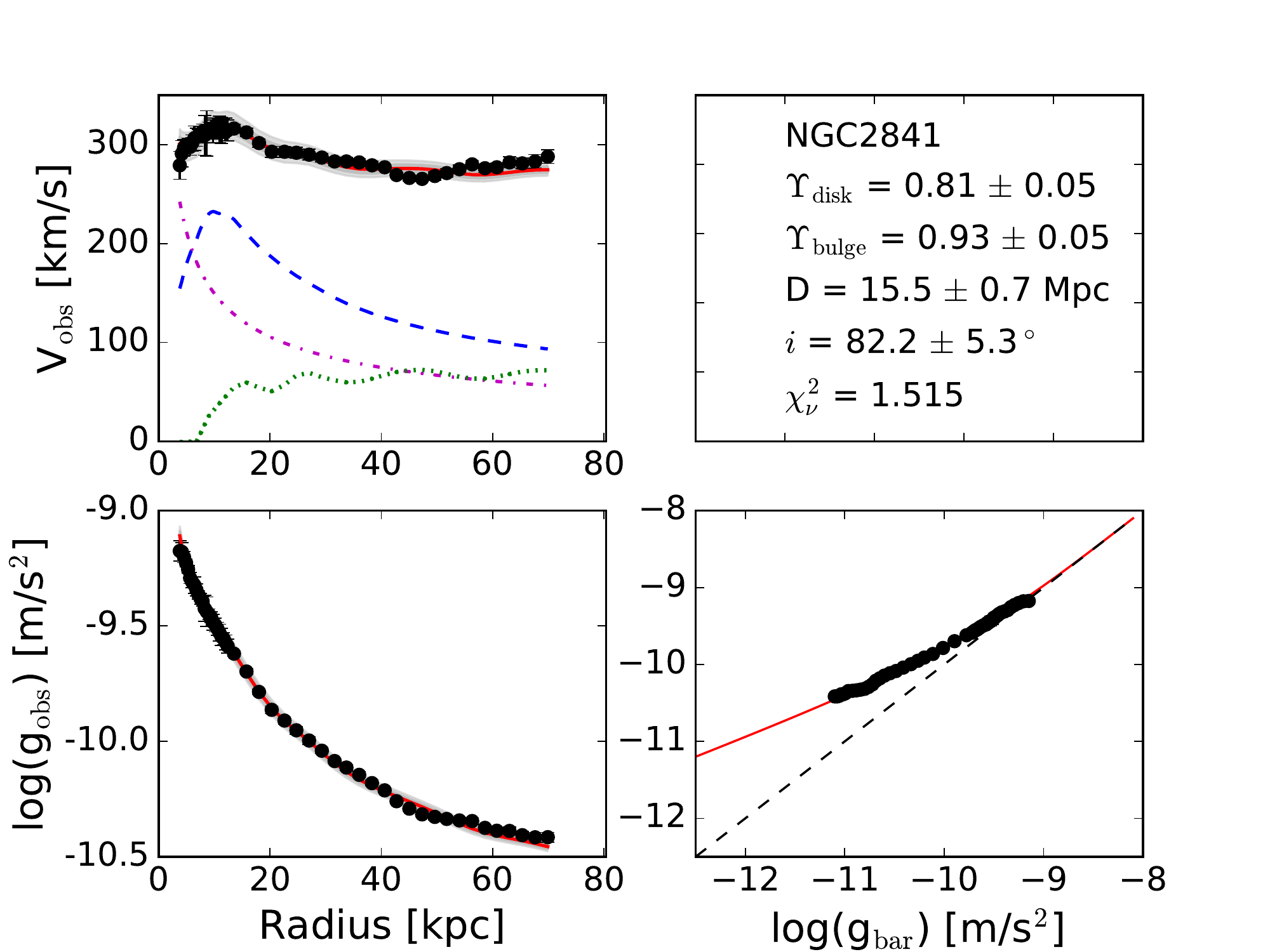}\includegraphics[scale=0.3]{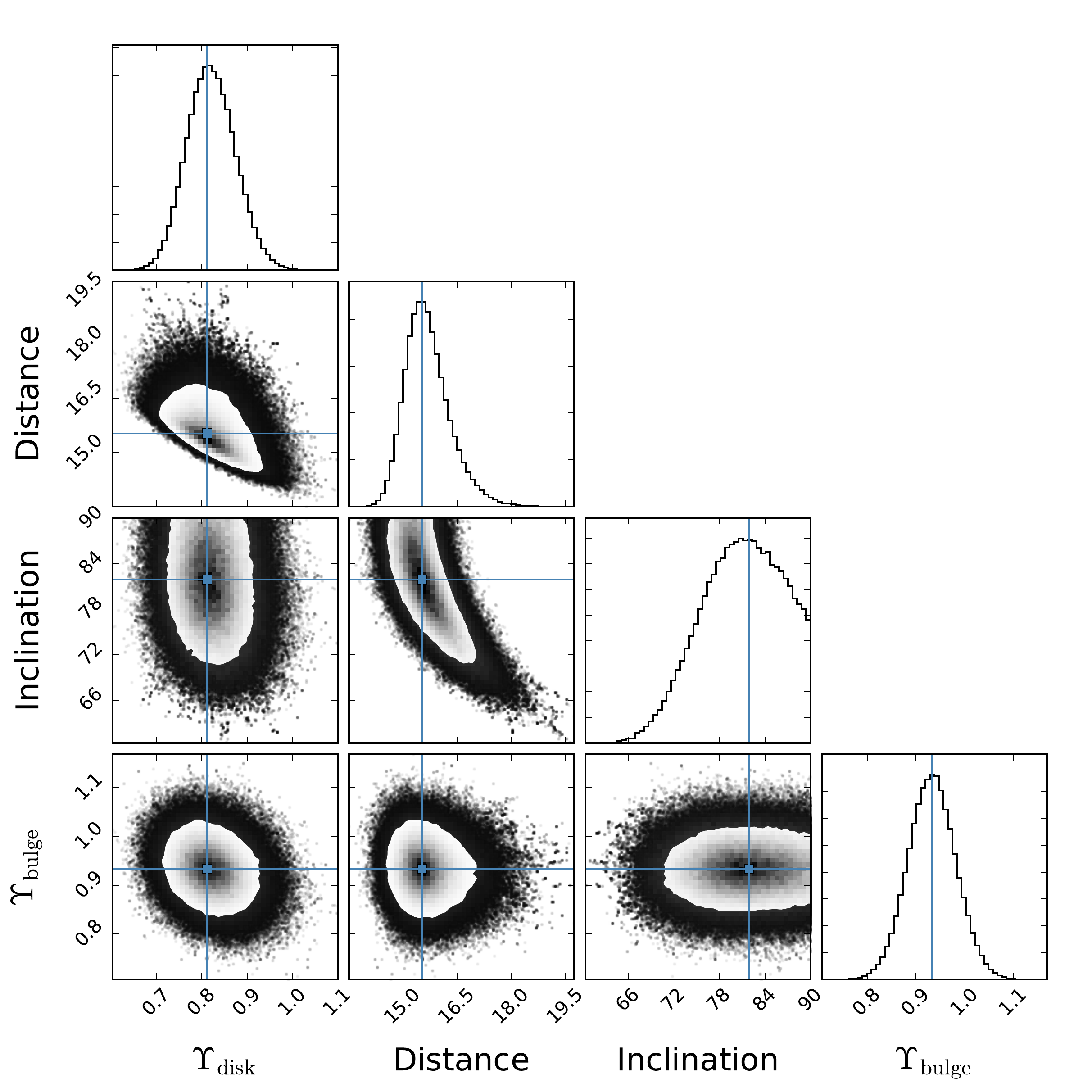}
\caption{Example of MCMC fits (\textit{left}) and the corresponding posterior distribution (\textit{right}). In the left panels, the points with error bars are the observed rotation curves $V_{\rm obs}(R)$ or corresponding accelerations $V^2_{\rm obs}(R)/R$. In the rotation curve panel, each baryonic component is presented: purple dash-dotted line for the bulge, blue dashed line for the disk, and green dotted line for the gas. The red solid line is the fitted rotation curve. The dark gray and light gray bands show the 68$\%$ and 95$\%$ confidence regions, respectively, considering the posterior distribution of $\Upsilon_{\star}$; they do not include additional uncertainties on $i$ and $D$. In the acceleration panels, the red solid line represents the mean RAR to which we fit. In the right panels, the blue cross indicates the parameter set corresponding to the maximum posterior probability. The complete figure set (175 images) is shown in the appendix.}
\label{N2841}
\end{figure*}

In section 2, we describe our fitting method, which is a Markov chain Monte Carlo (MCMC) simulation. In section 3, we show the fitted individual galaxies and their posterior distribution. The distributions of adjusted parameters are also presented. In section 4, the RAR and its residuals are described. We also check the resulting BTFR. We generalize the RAR to consider possible galaxy-to-galaxy variation in the critical acceleration scale in section 5. In section 6, we summarize our results and discuss the general implication of the extremely small rms scatter (0.057 dex) of our best-fitting relation.

\section{Method} \label{sec:method}

\subsection{Parameter dependence}

Based on Spitzer [3.6] images and total HI maps from the SPARC database, one can calculate the acceleration due to the baryonic mass distribution at every radius,
\begin{equation}
g_\mathrm{bar}(R) = (\Upsilon_\mathrm{disk}V^2_\mathrm{disk} + \Upsilon_\mathrm{bul}V^2_\mathrm{bul} + V^2_\mathrm{gas})/R,
\label{gbar}
\end{equation}
where $\mathrm{\Upsilon_{disk}}$ and $\mathrm{\Upsilon_{bul}}$ are the stellar mass-to-light ratios for the disk and bulge, respectively. Similarly, the observed acceleration can be calculated directly from the observed velocity $V_\mathrm{obs}$,
\begin{equation}
\centering
g_\mathrm{obs}(R) = \frac{V^2_\mathrm{obs}}{R}.
\end{equation}
According to the RAR \citep{McGaugh2014, 2016PhRvL.117t1101M, 2017ApJ...836..152L}, the expected total acceleration ${g_\mathrm{tot}}$ strongly correlates with that expected from baryonic distributions ${g_\mathrm{bar}}$,
\begin{equation}
g_\mathrm{tot}(R) = \frac{g_\mathrm{bar}}{1 - e^{-\sqrt{g_\mathrm{bar}/g_{\dagger}}}},
\label{mond}
\end{equation}
where $g_\dagger = 1.20 \times 10^{-10}$ m~s$^{-2}$.
Thus, one can compare the observed acceleration with the expected one. 

A constant value of $\Upsilon_\star$ for all galaxies is able to statistically establish the RAR, but some scatter must have been introduced since $\Upsilon_\star$ should vary from galaxy to galaxy. An inappropriate $\Upsilon_\star$ can lead to systematic offsets from the RAR for individual objects. Specifically, $\Upsilon_{\rm disk}$ and $\Upsilon_{\rm bulge}$ affect ${g_\mathrm{bar}}$ according to Equation \ref{gbar}.

Uncertainties in galaxy distance affect the radius ($R$) and the baryonic components of the rotation curve ($V_{\rm k}$). With $D$ being adjusted to $D'$, $R$ and $V_{\rm k}$ transform as

\begin{equation}
\centering
R' = R\frac{D'}{D};\ \ V'_{k} = V_{k}\sqrt{\frac{D'}{D}}
\label{changedis}
\end{equation}
where $k$ denotes disk, bulge or gas. Therefore, ${g_\mathrm{bar}}$ does not depend on distance. Instead, ${g_\mathrm{obs}}$ goes as $D^{-1}$ because the observed rotation velocity ($V_{\rm obs}$) and its error ($\delta V_{\rm obs}$) are inferred from the line-of-sight velocity which is distance independent. 

In the SPARC database, galaxy distances are estimated using five different methods (see \citealt{2016AJ....152..157L} for details): (1) the Hubble flow corrected for Virgo-centric infall (97 galaxies), (2) the tip of the red giant branch (TRGB) method (45 galaxies), (3) the magnitude-period relation of Cepheids (3 galaxies), (4) membership to the Ursa Major cluster of galaxies (28 galaxies), and (5) supernovae (SN) light curves (2 galaxies). The first method is the least accurate because the systemic velocity (redshift) of a galaxy may be largely affected by peculiar flows in the nearby Universe. The other methods have accuracies ranging between 5\% and 15\%. Table I in \citet{2016PhRvL.117t1101M} shows that errors on galaxy distance are the main source of scatter on the RAR. 

\begin{figure*}[t]
\centering
\includegraphics[scale=0.52]{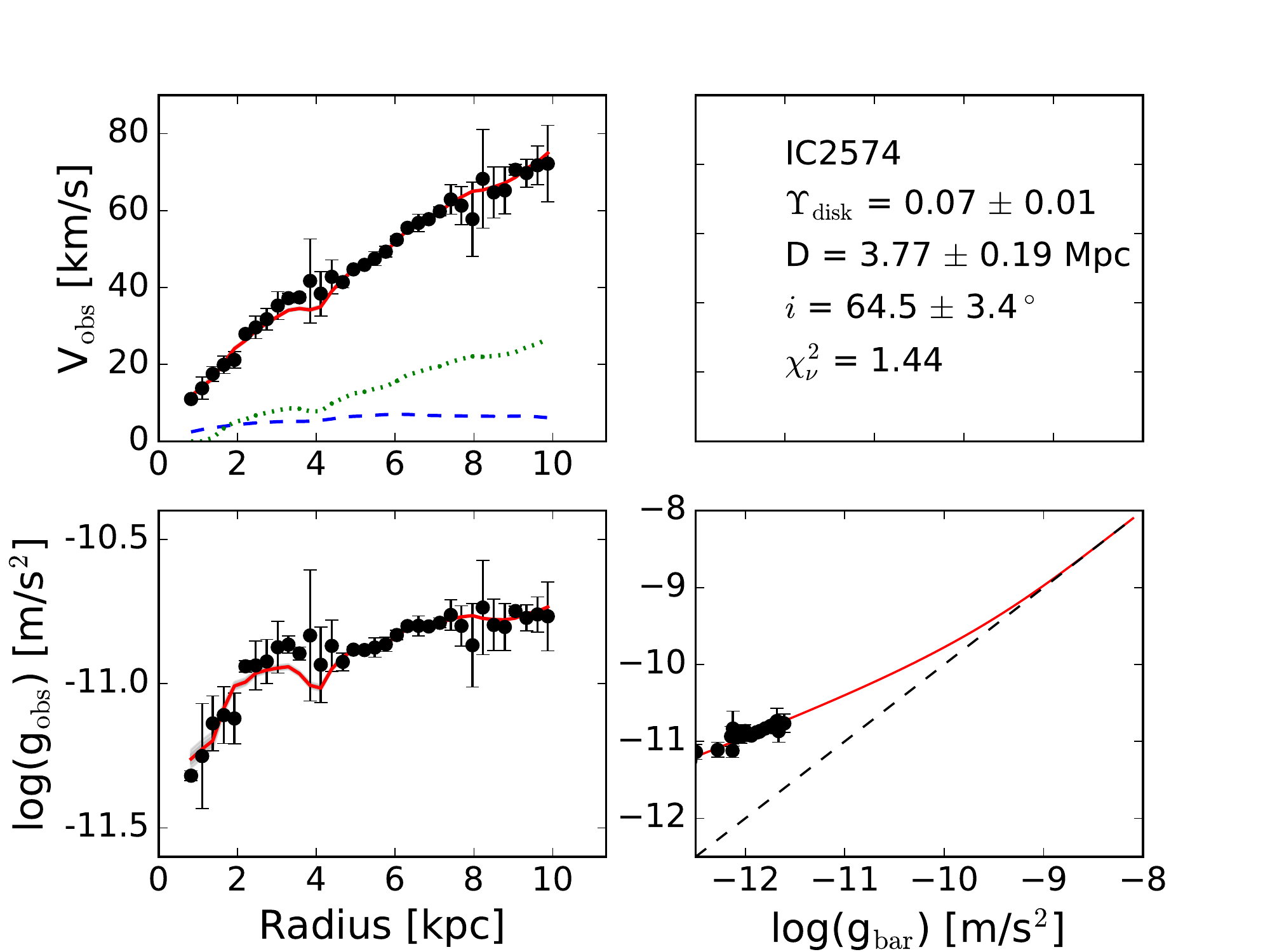}\includegraphics[scale=0.39]{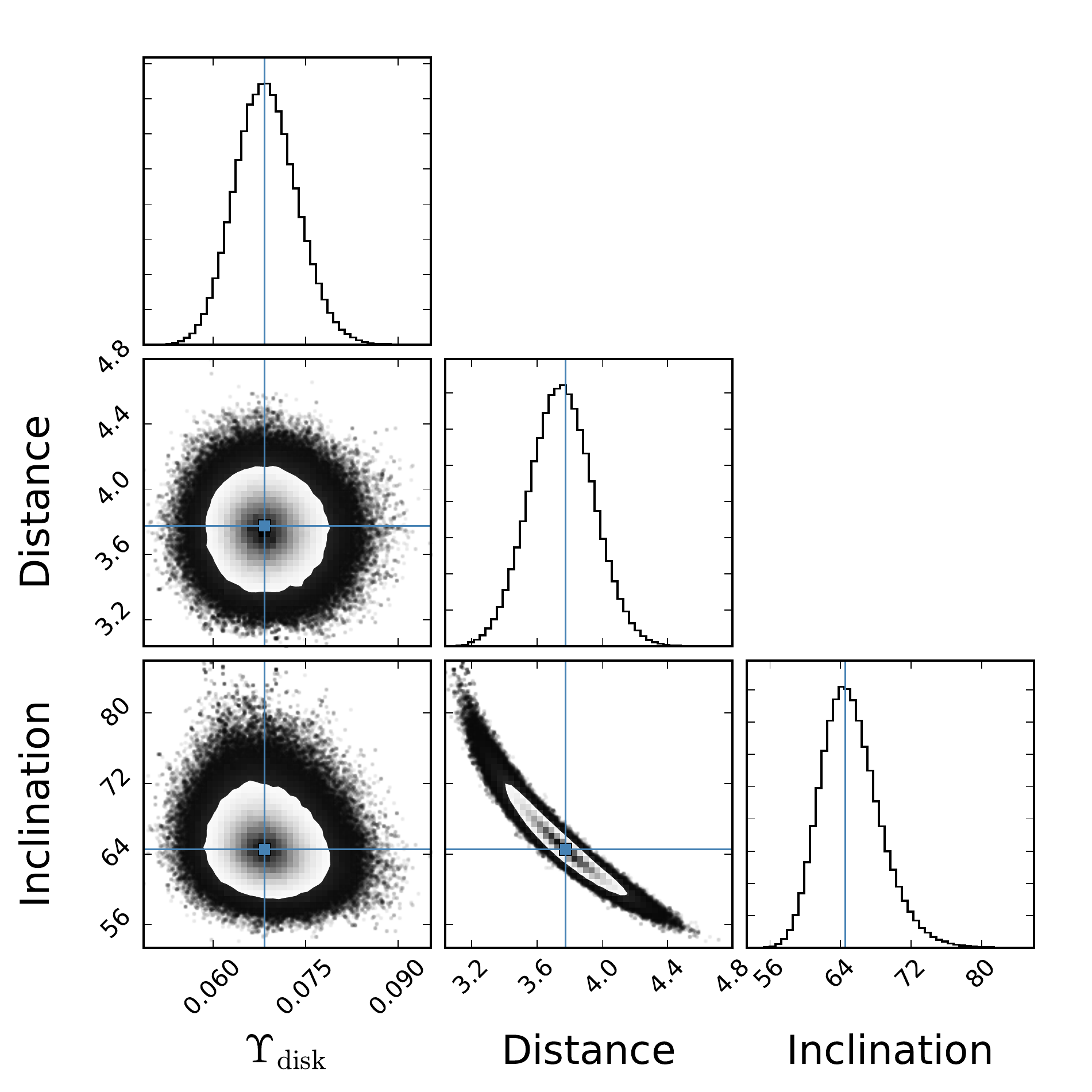}
\caption{Same as Figure\,\ref{IC2574} but for the gas-dominated dwarf galaxy IC\,2574.}
\label{IC2574}
\end{figure*}

Uncertainties in disk inclination are another important source of scatter. When the disk inclination $i$ is adjusted to $i'$, $V_{\rm obs}$ and $\delta V_{\rm obs}$ transform as
\begin{equation}
V'_{\rm obs} = V_{\rm obs}\frac{\sin(i)}{\sin(i')};\ \  \delta V'_{\rm obs} = \delta V_{\rm obs}\frac{\sin(i)}{\sin(i')}.
\label{changeinc}
\end{equation}
Hence, ${g_\mathrm{obs}}$ has a further dependence on disk inclination. Clearly, the correction becomes very large for face-on galaxies with small inclination. We also note that several galaxies have warped HI disks: the inclination angle systematically varies with radius. While warps are taken into account in deriving rotation curves (e.g., NGC 5055 from \citealt{2005MNRAS.364..433B}), here we treat the inclination as a single global parameter for each galaxy.

\subsection{MCMC simulation}\label{MCMC}

To fit individual galaxies, we used $emcee$ \citep{2013PASP..125..306F} to map the posterior distribution of the parameter set: the stellar mass-to-light ratio, galaxy distance, and disk inclination. Following standard procedures, we imposed Gaussian priors on $\Upsilon_\star$, $D$, and $i$. The priors were centered around the assumed values in SPARC and have standard deviations given by the observational errors for $D$ and $i$ and the scatter expected from stellar population models for $\Upsilon_{\star}$ \citep[e.g.,][]{Bell2001}. Hence, $\Upsilon_\star$, $D$, and $i$ are not entirely free parameters: the MCMC simulation is searching for an optimal solution within a realistic region of the parameter space. Specifically, we imposed $\Upsilon_{\rm disk}=0.5$ and $\Upsilon_{\rm bulge}=0.7 \;\mathrm{M}_{\sun}/\mathrm{L}_{\sun}$ with a standard deviation of 0.1 dex. We adopted a fixed mass-to-light conversion for the gas unlike what \citet{Swaters2012} did. We also required that the parameters remain physical and positive definite: $\Upsilon_{\star} > 0$,
$D > 0\;\mathrm{Mpc}$, and $0^{\circ} < i < 90^{\circ}$.

We used the standard affine-invariant ensemble sampler in $emcee$ and initialized the MCMC chains with 200 random walkers. We ran 500 burnt-in iterations and then ran the simulation to more than five autocorrelation times. We checked that the acceptance fractions for all galaxies are in the range (0.1, 0.7). To achieve the acceptance fraction, we set the size of the stretch move $a$ = 2.

We record the parameter set corresponding to the maximum probability and calculate the reduced $\chi^2$, 
\begin{equation}
\chi^2_\nu = \sum_R\frac{[g_\mathrm{obs}(R) - g_\mathrm{tot}(R)]^2/\sigma^2_{g_\mathrm{obs}}}{N - f},
\end{equation}
where $\sigma_{g_{\rm obs}} = 2 V_{\rm obs}\times\frac{\delta V_{\rm obs}}{R}$ is the uncertainty in the observed acceleration, $N$ the number of data points, and $f$ the degrees of freedom, for every galaxy.

\begin{figure*}[t]
\includegraphics[width=0.5\hsize]{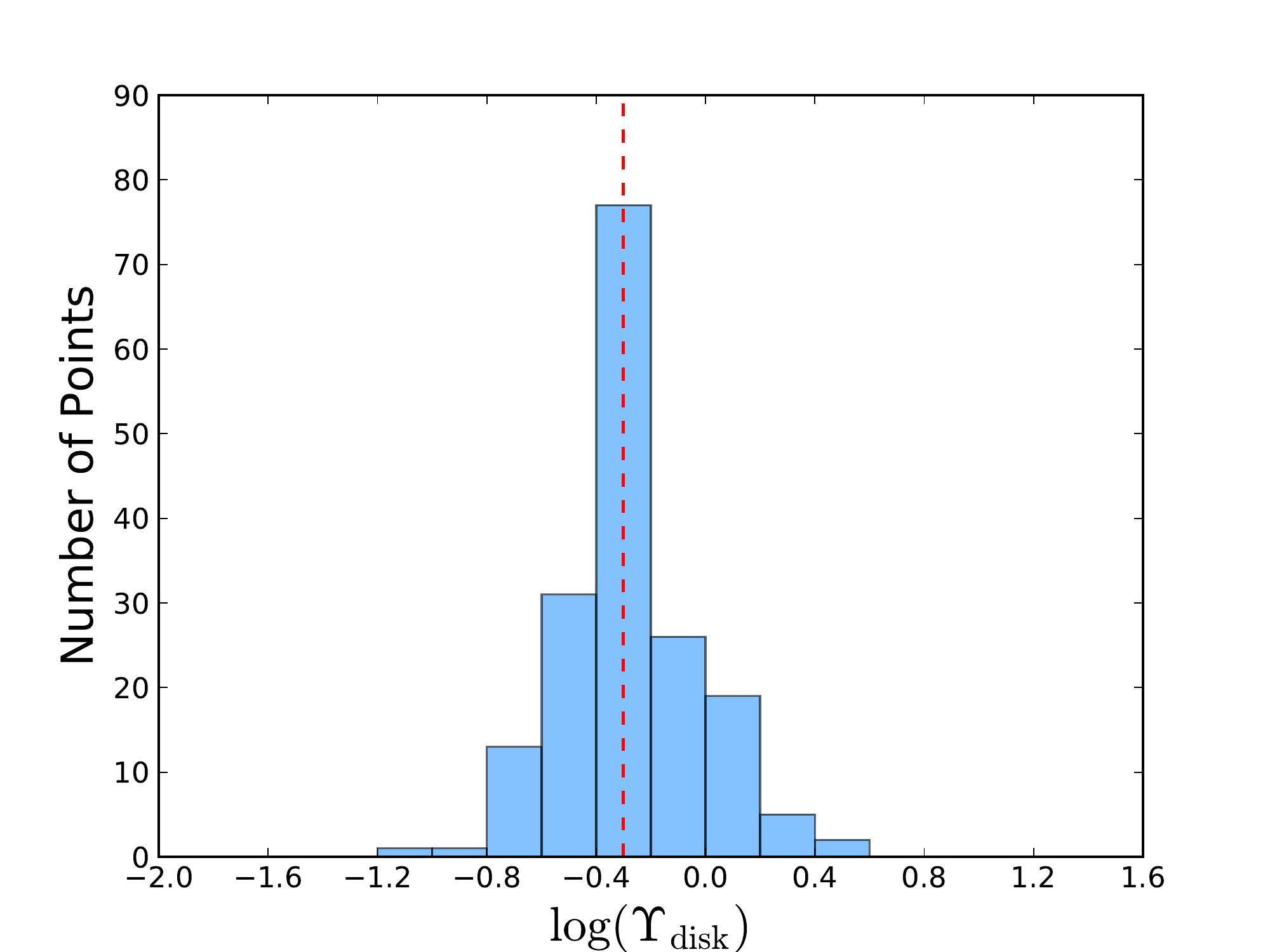}\includegraphics[width=0.5\hsize]{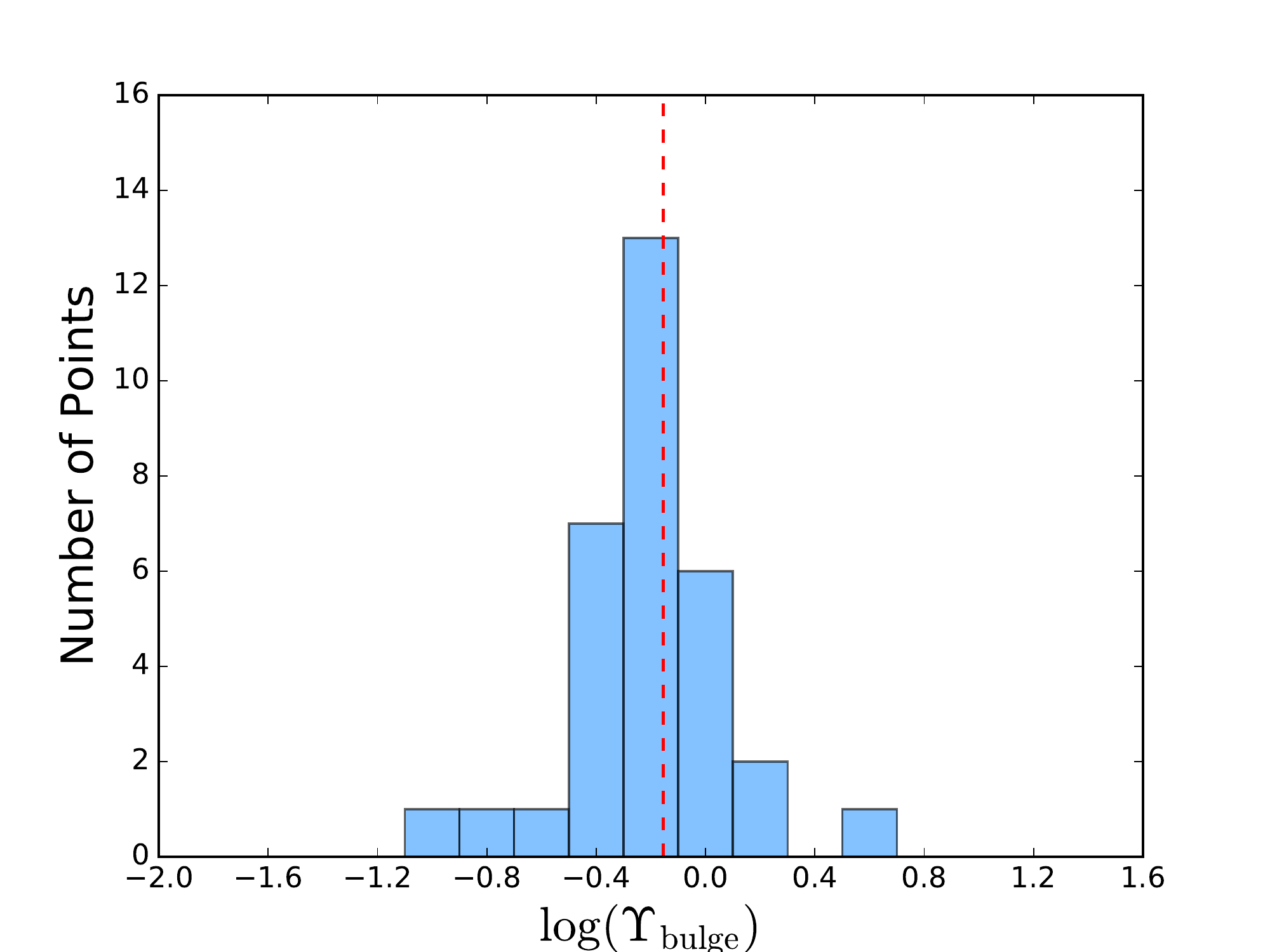}
\includegraphics[width=0.5\hsize]{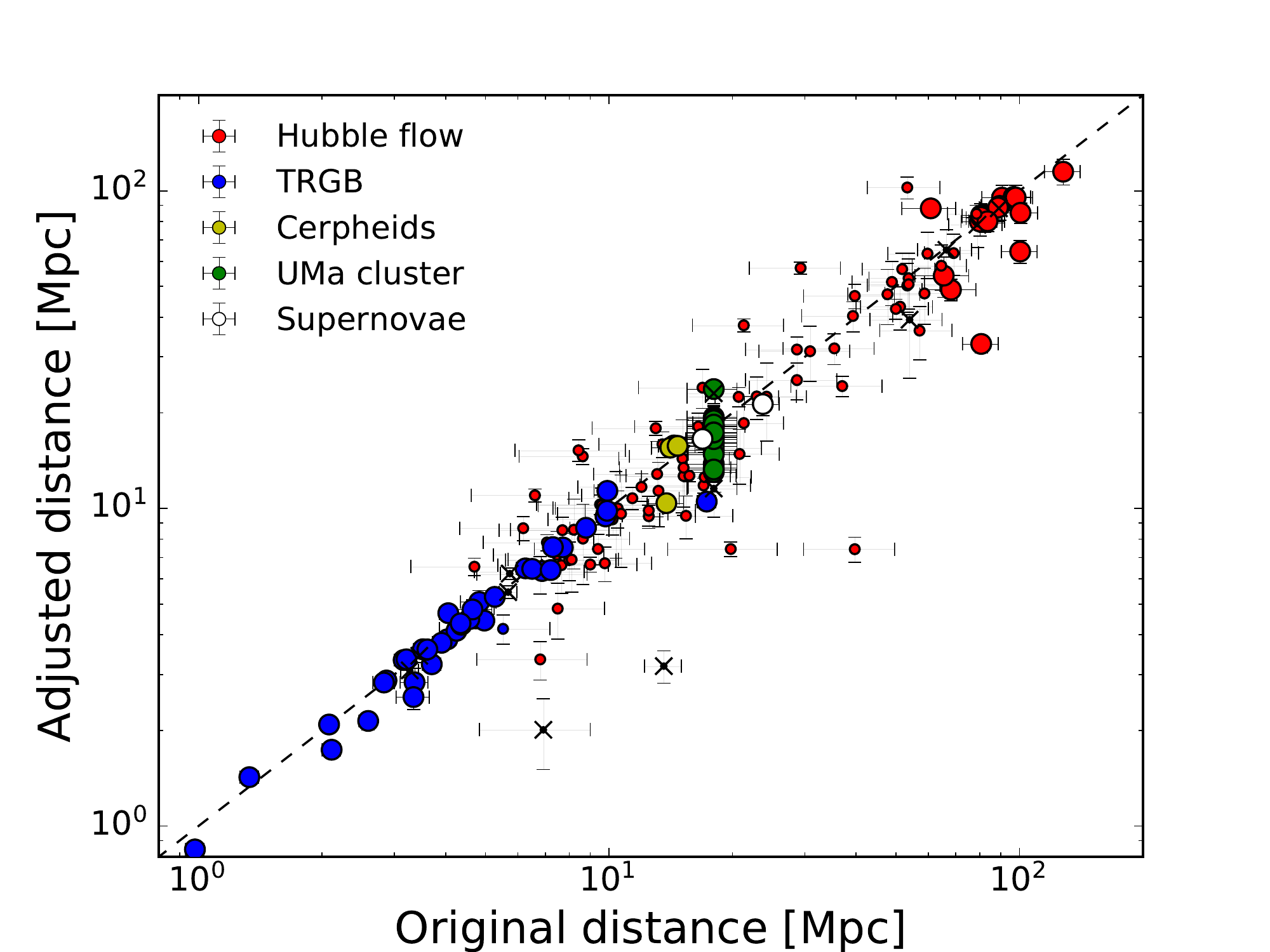}\includegraphics[width=0.5\hsize]{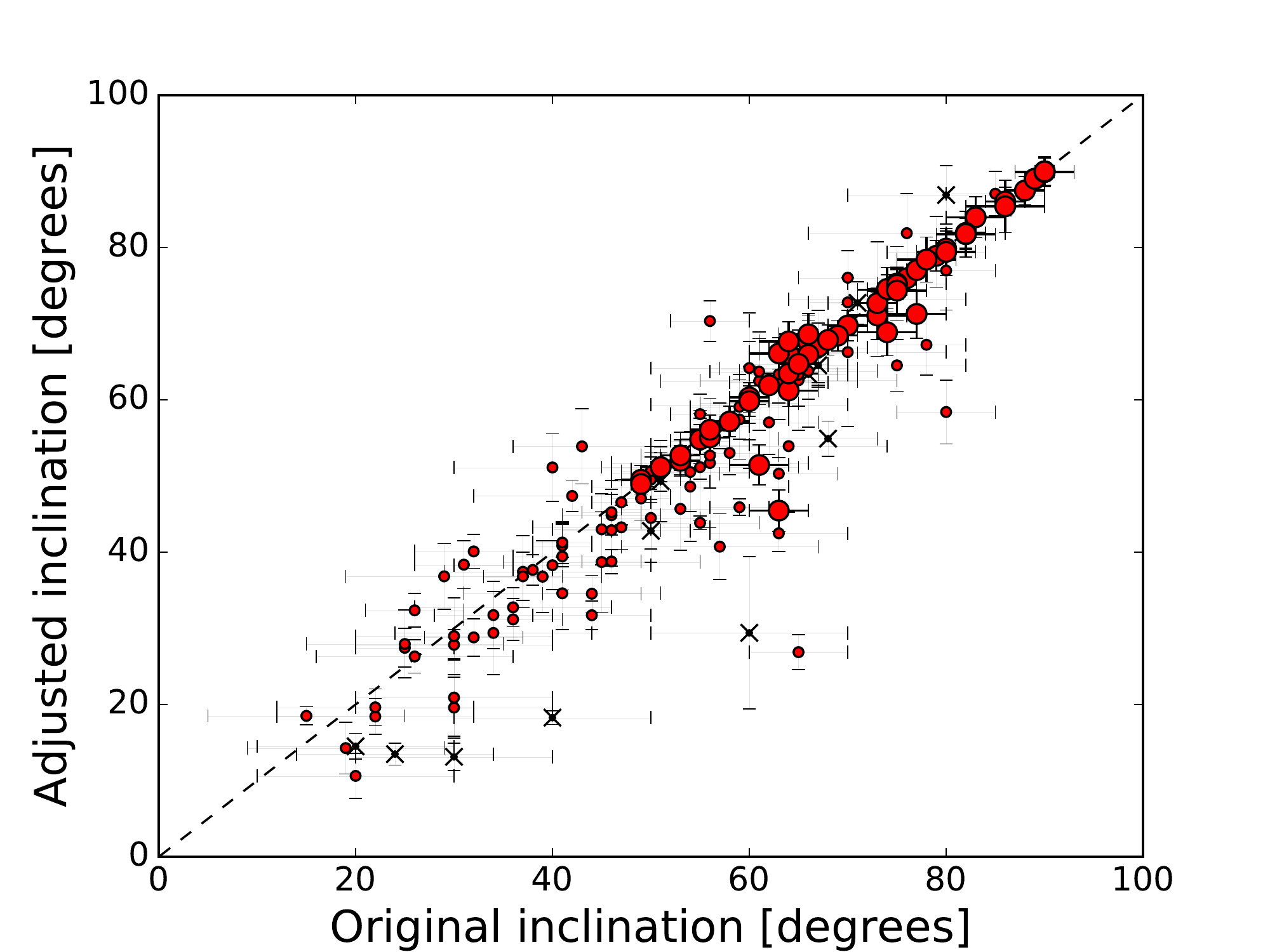}
\caption{Distributions of fitted parameters. The top panels show the histograms of the optimized $\Upsilon_{\rm disk}$ (top left) and $\Upsilon_{\rm bulge}$ (top right). The vertical dashed lines represent the values of 0.5 (disk) and 0.7 (bulge) adopted in \citet{2016PhRvL.117t1101M}. In the bottom panels, the optimal galaxy distance (bottom left) and disk inclination (bottom right) are plotted against their original values. The dashed line is the line of unity. Different methods of measuring distance are indicated by different colors. Large and small symbols correspond to data with an accuracy higher and lower than 15\% for distance and 5\% for inclination, respectively, based on observational errors as tabulated in SPARC. Crosses indicate galaxies with the low-quality rotation curves (Q=3, see \citealt{2016AJ....152..157L}). A few other outliers are discussed in the text.}
\label{parameter}
\end{figure*}

\section{Fitting individual galaxies}\label{sec:fit}

\subsection{Fit results}\label{sec:individual}

By fitting individual galaxies to the RAR, the stellar mass-to-light ratio, the galaxy distance and the disk inclination are optimized. In particular, we note that the RAR may be used as a distance indicator in analogy to the BTFR. The former relation is more demanding in terms of data quality, but has the advantage of using the full shape of the rotation curve and the baryonic mass profile instead of merely using the flat rotation velocity and total baryonic mass that go into the BTFR.

Figure\,\ref{N2841} shows an example of an MCMC fit for a star-dominated spiral galaxy (NGC\,2841). This object has historically been regarded as a problematic case for MOND \citep{Begeman1991, Gentile2011}, but a good fit is obtained allowing for uncertainties in distance and inclination within 1$\sigma$. The values of $\Upsilon_{\rm disk}$ and $\Upsilon_{\rm bulge}$ are relatively high but not unreasonable for such a massive, metal-rich galaxy. Similar figures are available for all SPARC galaxies.

Figure\,\ref{IC2574} illustrates the MCMC fit of a gas-dominated dwarf galaxy (IC\,2574). This object is often considered a problematic case for $\Lambda$CDM because it has a large core extending over $\sim$8 kpc \citep[e.g.,][]{Oman2015}. Moreover, \citet{Navarro2017} claimed that this galaxy strongly deviates from the RAR (their Figure 3). We find an excellent fit for IC~2574 after adjusting its distance and inclination by 1 $\sigma_D$ and 1.5 $\sigma_i$, respectively. The adjusted mass-to-light ratio is rather low ($\Upsilon_{\rm disk}=0.07$ M$_{\odot}$/L$_{\odot}$), perhaps uncomfortably so. This object is also present in the THINGS database \citep{deBlok2008}, where the  rotation curve is consistent with but slightly higher than that adopted here. If we apply the same MCMC technique to the THINGS data, we find a good fit with $\Upsilon_{\star} = 0.25$ M$_{\odot}$/L$_{\odot}$, illustrating the sensitivity of this parameter to even small changes in the rotation curve.

We stress that for gas-rich dwarfs, the vast majority of the rotation curve is explained by the gas contribution with very little room for adjustment. Rather than be overly concerned with the exact value of $\Upsilon_{\star}$ in such gas-dominated galaxies, it is amazing that this procedure works at all: $\Upsilon_{\star}$ has little power to affect the overall fit, while $D$ and $i$ are constrained by their priors to be consistent with the observed values. Gas-dominated galaxies are more prediction than fit: given the observed gas distribution and the RAR, the rotation curve must be what it is. The fitting parameters provide only minor tweaks to the basic prediction.

In general, the fits to most galaxies are good. The mass-to-light ratios are generally consistent with the expectations of stellar population synthesis. It is rare that either $D$ or $i$ are adjusted outside of their observational uncertainties. We maintain the same fitting function (equation \ref{mond}) for all 175 galaxies.

While most fits are visually good, they may occasionally have poor values of $\chi^2$. These usually occur when one or a few individual velocity measurements have tiny error bars. The discrepancy of these points from the fit is small in an absolute sense, but still impacts $\chi^2$. It is likely that in some cases the errors are slightly underestimated. For example, the potential contribution of non-circular motions may have been understated and the velocities may not exactly trace the underlying gravitational potential. In general, these fits are as good as possible: one cannot do better with a dark matter halo fit. The Navarro-Frenk-White (NFW) halo fit to  NGC\,2841 \citep{Katz2017} looks indistinguishable from that in Fig.\ \ref{N2841}: the two extra fit parameters available with a dark matter fit do not alter the shape of the continuous line that best approximates the data. We therefore consider fits of this type to be good even if $\chi^2$ is larger than unity.

\begin{figure*}[t]
\includegraphics[width=0.5\hsize]{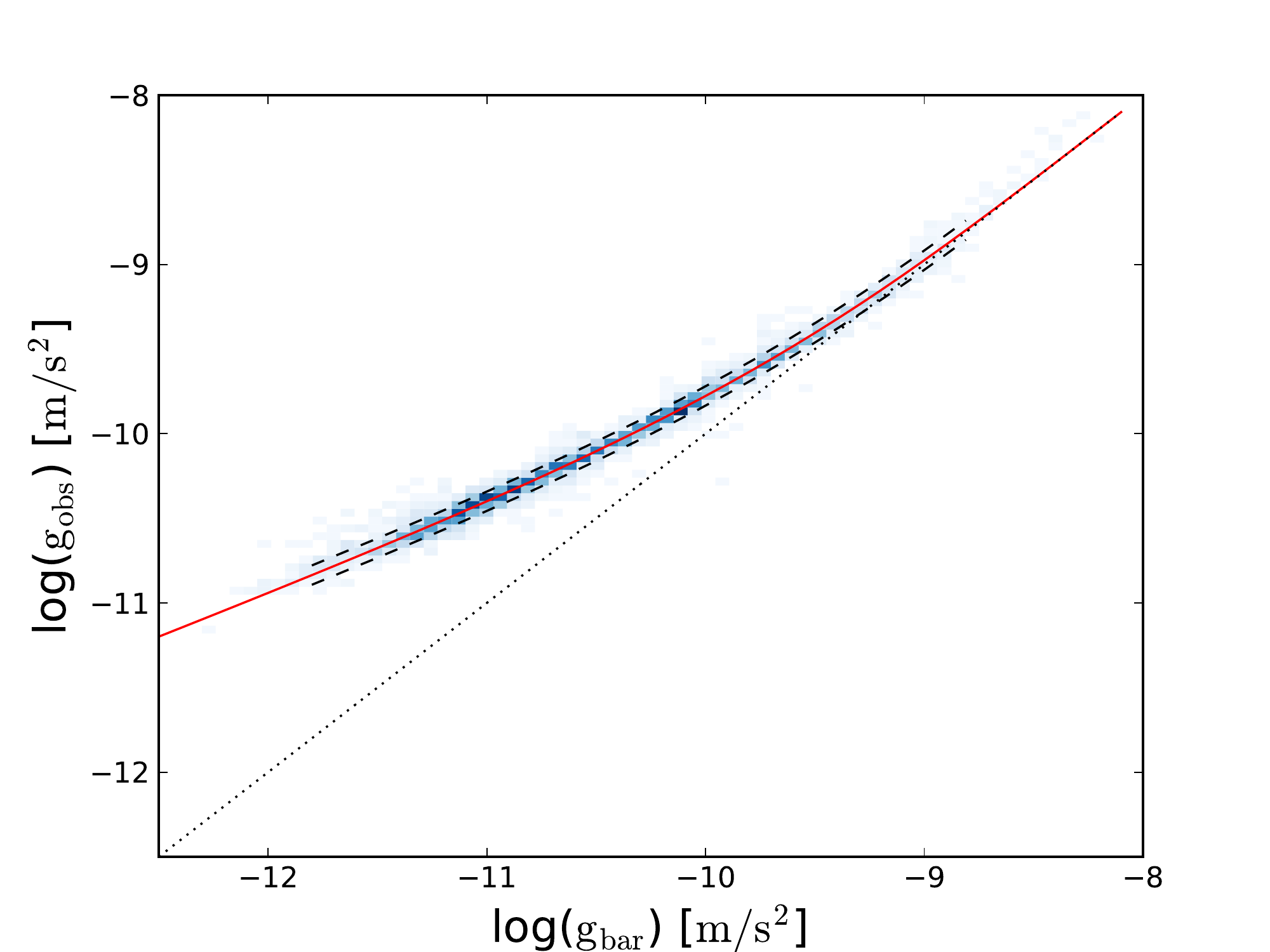}\includegraphics[width=0.5\hsize]{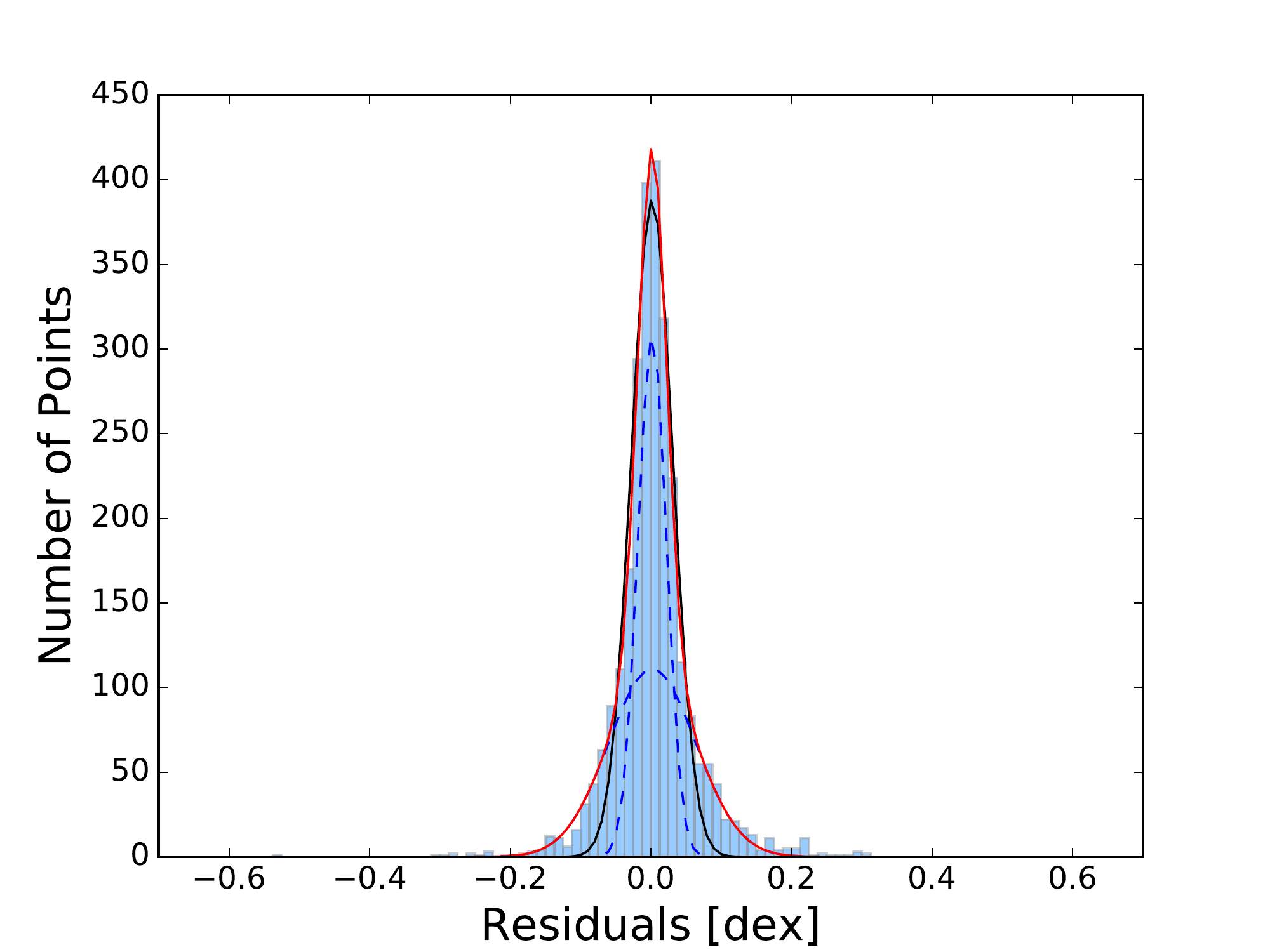}
\caption{RAR ($left$) and the residuals ($right$) with $\Upsilon_\star$, D, and $i$ optimized by the MCMC method. In the left panel, the red solid line represents the mean RAR from Equation \ref{mond}. The black dotted line is the line of unity. 2694 individual data points from 153 SPARC galaxies are represented by the blue color-scale. Black dashed lines show the rms scatter. In the right panel, the histogram of the residuals is fit with both single (black solid line) and double (red solid line) Gaussian functions. Blue dashed lines show the two components of the double Gaussian function.}
\label{RAR}
\end{figure*}

In about 10$\%$ of the cases, however, the fits are genuinely poor. Poor fits generally happen for rotation curves of lowest quality. For the sake of completeness, we fit all 175 galaxies in the SPARC database, but we did \textit{\textup{not}} expect to find good fits for galaxies with quality flag $Q=3$ (e.g., NGC\,4389, UGC\,2455, UGC\,4305) where the gas kinematics is likely out of equilibrium \citep[see][for details]{Lelli2016}. Some poor fits are also found for galaxies with $Q=1$ (e.g., D631-7, F571-8, and IC\,4202) and $Q=2$ (e.g., Cam\,B, DDO\,168, and NGC~2915). This may happen for several reasons: (i) the errors on $D$ and $i$ may be slightly underestimated, hence the priors place strong constraints that then contribute more to the probability function and preclude somewhat better fits, (ii) there may be features in the rotation curves that do not trace the smooth gravitational potential but are due to large-scale non-circular motions, and (iii) these galaxies may have unusual dust content that affects the shape of the [3.6] luminosity profile and the calculation of $g_{\rm bar}$ (this may be particularly important for edge-on systems such as IC\,4202 and F571-8). In the specific case of D631-7, the gas contribution was computed assuming a purely exponential distribution since the HI surface density profile was not available (see Lelli et al. 2016). Since the RAR is very sensitive to the precise baryonic distribution, even small deviations from an exponential profile may lead to a poor fit in such a gas-dominated galaxy. In general, we consider it likely that lower quality data lead to lower quality fits. Having $\sim$10$\%$ of such cases seems an inevitable occurrence in any astronomical database built from many diverse rotation curve studies such as SPARC \citep{2016AJ....152..157L}. It would be strange if there were no such cases.

\subsection{Distributions of adjusted parameters}

Figure \ref{parameter} shows the distributions of the optimized parameters. The top panels show histograms of $\Upsilon_\star$. The dashed lines indicate $\Upsilon_{\rm disk} = 0.5$ $M_{\odot}/L_{\odot}$ and $\Upsilon_{\rm bulge} = 0.7$ $M_{\odot}/L_{\odot}$ adopted in \citet{2016PhRvL.117t1101M} and \citet{2017ApJ...836..152L}. The optimized stellar mass-to-light ratios are tightly distributed around these values. The median values of $\Upsilon_{\rm disk}$ and $\mathrm{\Upsilon_{bulge}}$ are 0.50 and 0.67, respectively. By and large, the best-fit stellar mass-to-light ratios are consistent with the expectations of stellar population synthesis models \citep{Schombert2014a,Meidt2014,McGaugh2014b,Norris2016}.

Adjusted distances and inclinations are also shown in Figure \ref{parameter} (bottom panels). Galaxies with a low-quality flag in SPARC \citep[Q=3, see][]{2016AJ....152..157L} typically prefer smaller values of $D$ and $i$ with respect to their original values (see black crosses in the figure). After removing these low-quality data, the distributions of $D$ and $i$ are fairly symmetric around the line of unity indicating that there are no major systematics. Hubble flow distances are the least certain and show the largest variation between measured and best-fit distance. 
More accurate methods (Cepheids, TRGB) show less variation, as expected.

A few galaxies show significant deviations ($>$ 1 $\sigma_D$) from the optimized ones. For example, PGC51017 (with a TRGB distance) is a starburst dwarf galaxy where the rotation curve clearly does not trace the equilibrium gravitational potential \citep{2014A&A...566A..71L}, thus it is not surprising that the distance is pushed to unphysical values in order to obtain a good fit. This rotation curve has a low-quality flag in SPARC (Q=3) and is only included here for the sake of completeness. Another example is NGC3198, which is sometimes regarded as a problematic case for MOND \citep{Gentile2011, Gentile2013}. The MCMC method finds a good fit with $D=10.4\pm0.4$ Mpc, which is consistent with the Cepheid-based distance (13.8$\pm$1.4 Mpc) within 2$\sigma$.

Table in the appendix lists the optimal parameters and reduced $\chi^2$ for each galaxy in order of declining luminosity. The errors on the fitting parameters are estimated from their posterior distributions using the ``std'' output in GetDist. These errors are generally smaller than those in the SPARC database because
of the combined constraints from the Gaussian priors and the likelihood function.

\begin{table*}
\caption{BTFR: The fitted parameters and scatter.}
\label{table:BTFR}
\centering
\begin{tabular}{c|ccccc}
\hline\hline
{Case} & {Slope (n)} & {log(A)} & {Intrinsic scatter} & {rms scatter}\\
\hline
 Constant $\Upsilon_\star$ &3.81 $\pm$ 0.08 & 2.17 $\pm$ 0.17 & 0.108 $\pm$ 0.024 & 0.234 \\
 \hline
Free $\Upsilon_\star$, D, $i$ &3.79 $\pm$ 0.05 & 2.08 $\pm$ 0.10 & 0.035 $\pm$ 0.019 & 0.127 \\
\hline
\end{tabular}
\end{table*}
%---------------------------------------------------------------

\section{Galaxy scaling relations}

\subsection{Radial acceleration relation}\label{sec:RAR}

The RAR and its residuals, $\log(g_{\rm obs}) - \log(g_{\rm tot})$, are plotted in Figure \ref{RAR}. To compare them with previous results \citep{2016PhRvL.117t1101M, 2017ApJ...836..152L}, the same selection criteria were adopted: we removed 10 face-on galaxies with $i < 30^\circ$ and 12 galaxies with asymmetric rotation curves that do not trace the equilibrium gravitational potential (Q = 3). We also required a minimum precision of 10\% in observational velocity ($\delta V_{\rm obs}/V_{\rm obs} < 0.1$). This retains 2694 data points out of 3163.

After $\Upsilon_\star$, D, and $i$ were adjusted within the errors, the RAR was extremely tight and had an rms scatter of 0.057 dex. We fit the histogram of the residuals with a Gaussian function (the dashed line in the right panel of Figure \ref{RAR}): the fit is acceptable, but there are broad symmetric wings in the residuals that are not captured by a single Gaussian function. Hence, we fit a double Gaussian function.
The double Gaussian function substantially improves the fit and fully describes the residual distribution with standard deviations of 0.062 dex and 0.020 dex. 
The mean values $\mu$ are consistent with zero.

Interestingly, the errors on the rotation velocities are not expected to be Gaussian because they are obtained by summing two different contributors (see \citealt{2009A&A...493..871S,2016AJ....152..157L}): the formal error from fitting the whole disk (driven by data quality and random non-circular motions) and the difference between velocities in the approaching and receding sides of the galaxy (representing global asymmetries and kinematic lopsidedness). The success of the double Gaussian fit suggests that the two Gaussian components perhaps probe these two different sources of errors and hence dominate the total residual scatter over all other possible error sources.

The small residual scatter leaves very little room for any intrinsic scatter because (1) the observational errors in the rotation velocities are not negligible, driving errors in $g_{\rm obs}$, and (2) there could be additional sources of errors in $g_{\rm bar}$ like the detailed 3D geometry of baryons and possible radial variations in $\Upsilon_\star$. Considering these error sources, the intrinsic scatter in the RAR must be smaller than 0.057 dex.

\subsection{Baryonic Tully-Fisher relation}

The BTFR relates the total baryonic masses of galaxies to their flat rotation velocity. In some sense, this is the asymptotic version of the RAR at large radii (see Sect.\,7.1 of \citealt{2017ApJ...836..152L} for details). For $R\rightarrow\infty$, g$_{\rm bar}$ becomes small and Eq.\ \ref{mond} gives $g_{\rm obs} \simeq \sqrt{g_{\rm bar}g_\dagger}$ by Taylor expansion. Since $g_{\rm obs} = V_{\rm f}^2/R$ and $g_{\rm bar} \simeq GM_{\rm bar}/R^2$, the radial dependence cancels out and we are left with $V^4_{\rm f}\propto M_{\rm bar}$. Thus, a BTFR with slope 4 is built into eq.\ \ref{mond}.

Here, we fit the BTFR directly to check how well we recover the behavior required by eq.\ \ref{mond}. In addition to the slope specified by the asymptotic limit of the RAR, we also expect the BTFR to be tighter after fitting $\Upsilon_\star$, $D,$ and $i$ to the RAR. However, this does not necessarily have to happen since the BTFR only considers the flat rotation velocity ($V_\mathrm{f}$) and the total baryonic mass ($M_b$), whereas we fit the whole shape of the rotation curve using the full baryonic mass profile. We adopted the same selection criteria as described in Sect.\,\ref{sec:RAR} and removed the galaxies that did not reach a flat rotation velocity as defined by \citet{Lelli2016}. This retained 123 galaxies out of 175 (5 more galaxies with the latest version of SPARC relative to \citealt{Lelli2016}). 

\begin{figure}[t]
\includegraphics[width=\hsize]{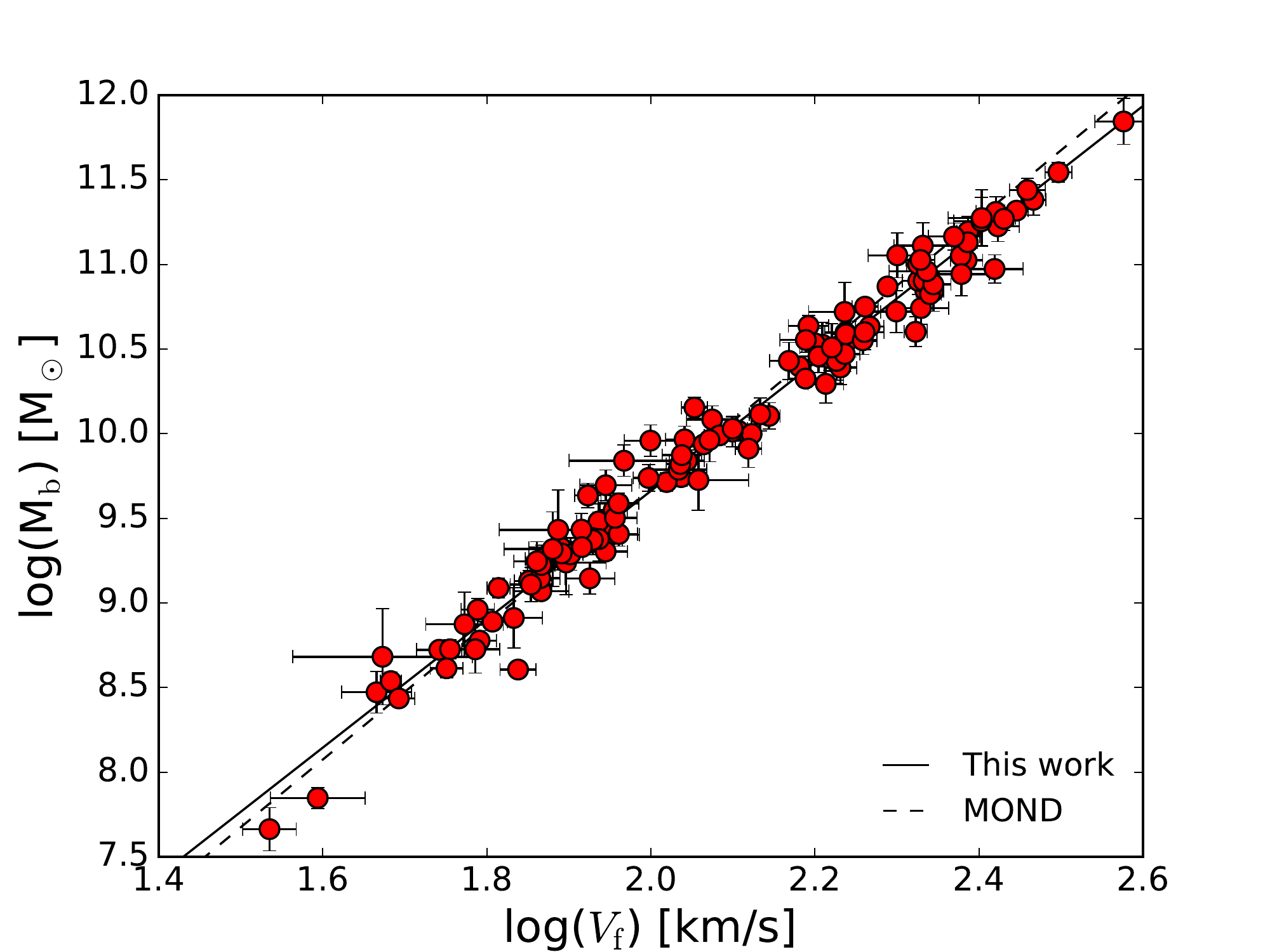}
\caption{Baryonic Tully-Fisher relation with $\Upsilon_\star$, D, and $i$ optimized (red points). The solid line illustrates the fitted BTFR. For reference, the dashed line shows the prediction of MOND: $\log(M_\mathrm{b}) = 4\log(V_\mathrm{f}) + \log[X/(a_0 G)]$.}
\label{BTFR}
\end{figure}

Figure \ref{BTFR} shows a tight BTFR. We used the LTS\_LINEFIT program \citep{2013MNRAS.432.1709C} to fit the linear relation
\begin{equation}
\log(M_\mathrm{b}) = n\log(V_\mathrm{f}) + \log(A).
\end{equation}
LTS\_LINEFIT considers errors in both variables and estimates $n$ and $A$ together with the intrinsic scatter around the linear relation.
The errors on $V_{\rm f}$ and $M_{\rm b}$ were calculated using equations 3 and 5 of \citet{Lelli2016}, but we treated disk and bulge separately. These equations consider the errors on $\Upsilon_{\rm disk}$, $\Upsilon_{\rm bulge}$, $i$, and $D$, which were estimated from the posterior distributions of the MCMC fits. We also corrected observed quantities such as luminosity and flat rotation velocity according to the adjusted $D$ and $i$. The fitting results are summarized in Table \ref{table:BTFR}. To enable a direct comparison with \citet{Lelli2016}, we also show the case where $D$ and $i$ were kept fixed to the SPARC values and the mass-to-light ratio was constant for all galaxies, but we improved compared to \citet{Lelli2016} by considering disk ($\Upsilon_{\star} = 0.5$) and bulge ($\Upsilon_{\star} = 0.7$) separately. In this case, the errors on $D$ and $i$ were taken from the SPARC database, while the errors on $\Upsilon_{\rm disk}$ and $\Upsilon_{\rm bulge}$ were assumed to be 0.11 dex for all galaxies. We find a slightly steeper slope than \citet{Lelli2016} because bulges have higher mass-to-light ratios than disks and are more common in massive galaxies, increasing $M_{\rm b}$ at the top end of the relation. Except for this small difference, our results are entirely consistent with those of \citet{Lelli2016} for the constant $\Upsilon_{\star}$ case.

The estimated intrinsic scatter of the BTFR is rather small. We estimate a conservative upper limit on the intrinsic scatter as the best-fit value (0.035 dex) plus $3 \sigma$ (the error on the estimated intrinsic scatter). This gives $\sigma_{\rm intr} < 0.1$ dex. Satisfying this bound provides a strong constraint on galaxy formation models (e.g., \citealt{Desmond2017b}).

The fitted slope of the BTFR is close to 3.8 in both cases. Formally, this differs from 4.0 by $\sim 4 \sigma$, although slopes consistent with 4 are obtained when we weight the data by the gas fraction \citep{Lelli2016}. Given that the functional form of Eq.\,\ref{mond} guarantees a BTFR with a slope of 4, this discrepancy is puzzling.

Several effects may cause the difference. One is simply that there are uncertainties in the mass-to-light ratios estimated from our fits. This adds scatter to the data, which inevitably lowers the fitted slope.

A more subtle concern is that the measured value of $V_{\rm f}$ is not quite the same as that implicit in Equation \ref{mond}. We measure $V_{\rm f}$ in the outer parts of extended rotation curves, and can do so consistently and robustly. However, Equation \ref{mond} only guarantees a BTFR slope of 4 with $V_{\rm f}$ in the limit of zero acceleration, or infinite distance from the galaxy. The measurements are made at finite radii. The definition of $V_{\rm f}$ in \citet{Lelli2016} (their equation 2) requires measured rotation curves to be flat to within 5\%, but there may be some small slope within that limit. It is well known (e.g., \citealt{2001ApJ...563..694V}) that bright galaxies have rotation curves that tend to decline toward $V_{\rm f}$ , while those of faint galaxies tend to rise toward $V_{\rm f}$. It is conceivable that this effect causes a slight systematic variation in the measured $V_{\rm f}$ with mass that acts to lower the slope. That is to say, the value of $V_f$ we measure empirically may not reach the flat velocity sufficiently well
that is implied by the limit $g_\mathrm{bar} \rightarrow 0$ assumed in the derivation above.

The geometry of disk galaxies may also have an impact: a thin disk rotates faster than the equivalent spherical distribution \citep{BT87}. This is quantified by the factor $X$ that appears in the normalization of the MOND prediction for the BTFR: $A = X/(a_0 G)$. The factor $X \rightarrow 1$ as $R \rightarrow \infty$, but on average, $\langle X \rangle = 0.8$ \citep{McGaugh2005} at the finite radii observed in spiral galaxies. We have assumed that $X$ is the same for all galaxies, but it is conceivable that disk thickness varies with mass so that $X$ is a weak function thereof. This might also affect the slope. 

Regardless of which of these effects dominates, it is clear that the slope of the BTFR is steep. It is not 3.0 as one might reasonably assume in $\Lambda$CDM (e.g., \citealt{1998MNRAS.295..319M}), nor is it 3.5 (e.g., \citealt{Bell2001}), as might be expected after adiabatic compression \citep{2001MNRAS.321..559B}. The difference (or lack thereof) between 3.8 and 4.0 may be the limit of what we can hope to discern with astronomical data. The limit is not due to the data themselves, but to systematic effects.

\section{Does the critical acceleration scale $g_{\dagger}$ vary?}

In the previous analysis we have assumed that $g_{\dagger}$ is constant for all galaxies. 
The small scatter observed around the RAR \citep{2016PhRvL.117t1101M,2017ApJ...836..152L} already demonstrates that this is very nearly the case.
However, the answer to the question of whether the value of $g_\dagger$ is truly constant can be used to distinguish between a scaling relation and a law of nature.

\begin{figure}[t]
\includegraphics[width=\hsize]{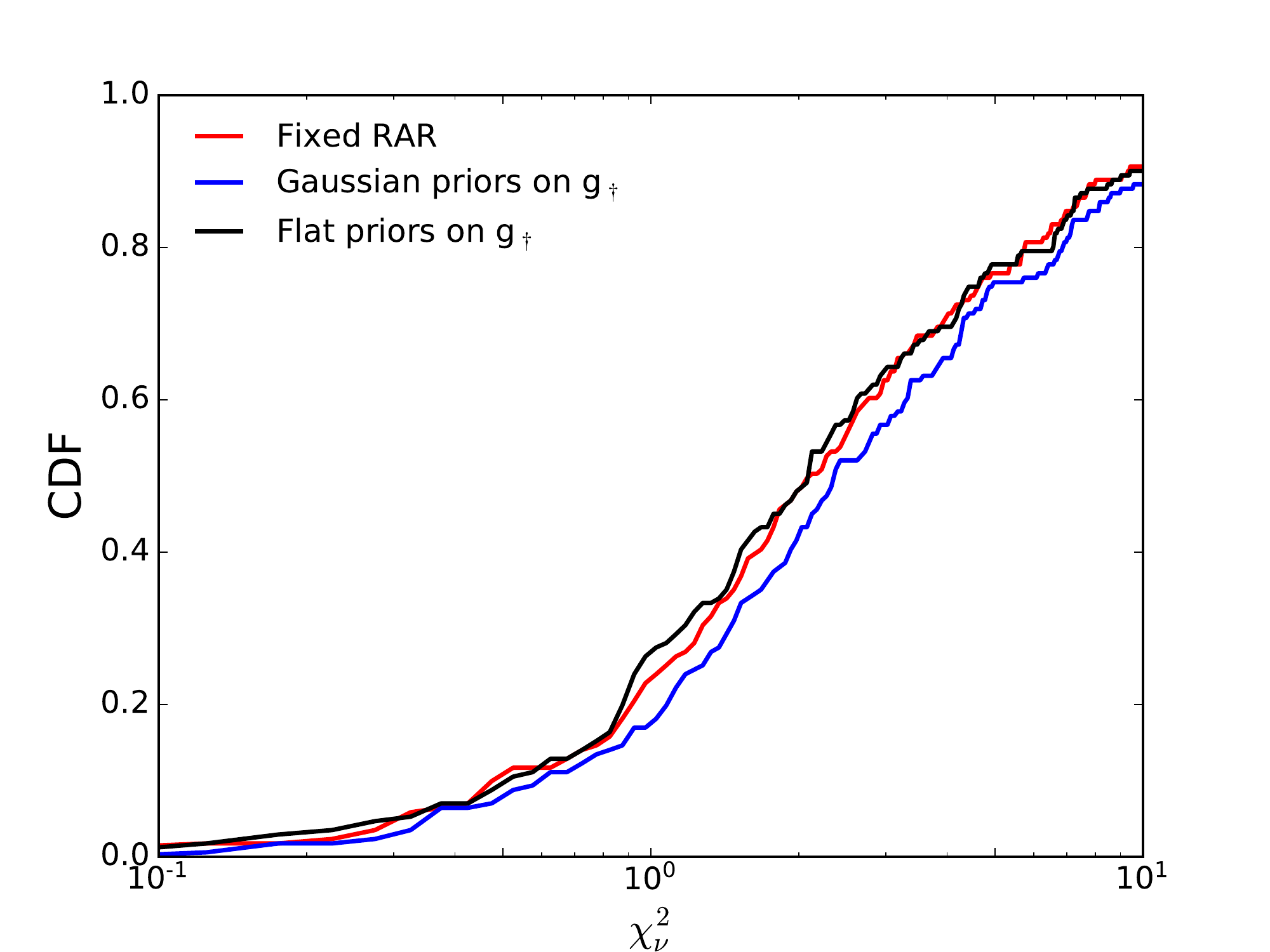}
\caption{Cumulative distributions of reduced $\chi^2$ for fixed $g_\dagger$ fits (red line) and variable $g_\dagger$ fits using flat (black line) and Gaussian priors (blue line).}
\label{CDF_Chi}
\end{figure}

As a further check on this point, we fit all galaxies again, treating $g_{\dagger}$ as an additional free parameter. 
We made fits with both a flat prior (for the range $0 \le g_{\dagger} \le 10^{-9}\;\mathrm{m}\,\mathrm{s}^{-2}$)  and a Gaussian prior
\citep[with $g_{\dagger} = 1.20\pm0.02\;\mathrm{m}\,\mathrm{s}^{-2}$: ][]{2016PhRvL.117t1101M,2017ApJ...836..152L}.
The cumulative distributions of reduced $\chi^2$ of these fits are shown along with that for fixed $g_{\dagger}$ in Figure \ref{CDF_Chi}.

Allowing $g_{\dagger}$ to vary from galaxy to galaxy does not improve the fits. 
In the case of the flat prior, the cumulative distribution of reduced $\chi^2$ is practically indistinguishable from the case of fixed $g_{\dagger}$,
despite the additional freedom.
In the case of a Gaussian prior, the reduced $\chi^2$ is even
slightly worse because essentially the same fit is recovered 
(for a similar total $\chi^2$), but the extra parameter increases the number of degrees of freedom, increasing the reduced $\chi^2$. 
The fits are not meaningfully improved by allowing $g_{\dagger}$ to vary.

The resulting rms scatter remains nearly invariant: 0.054 dex and 0.057 dex when using the flat prior and the Gaussian prior on g$_\dagger$, respectively. This indicates that the remaining rms scatter is dominated by observational uncertainties on rotation curves and possible intrinsic scatter. As a practical matter, there is no room to accommodate substantial variation in g$_\dagger$.

\begin{figure}[t]
\includegraphics[width=\hsize]{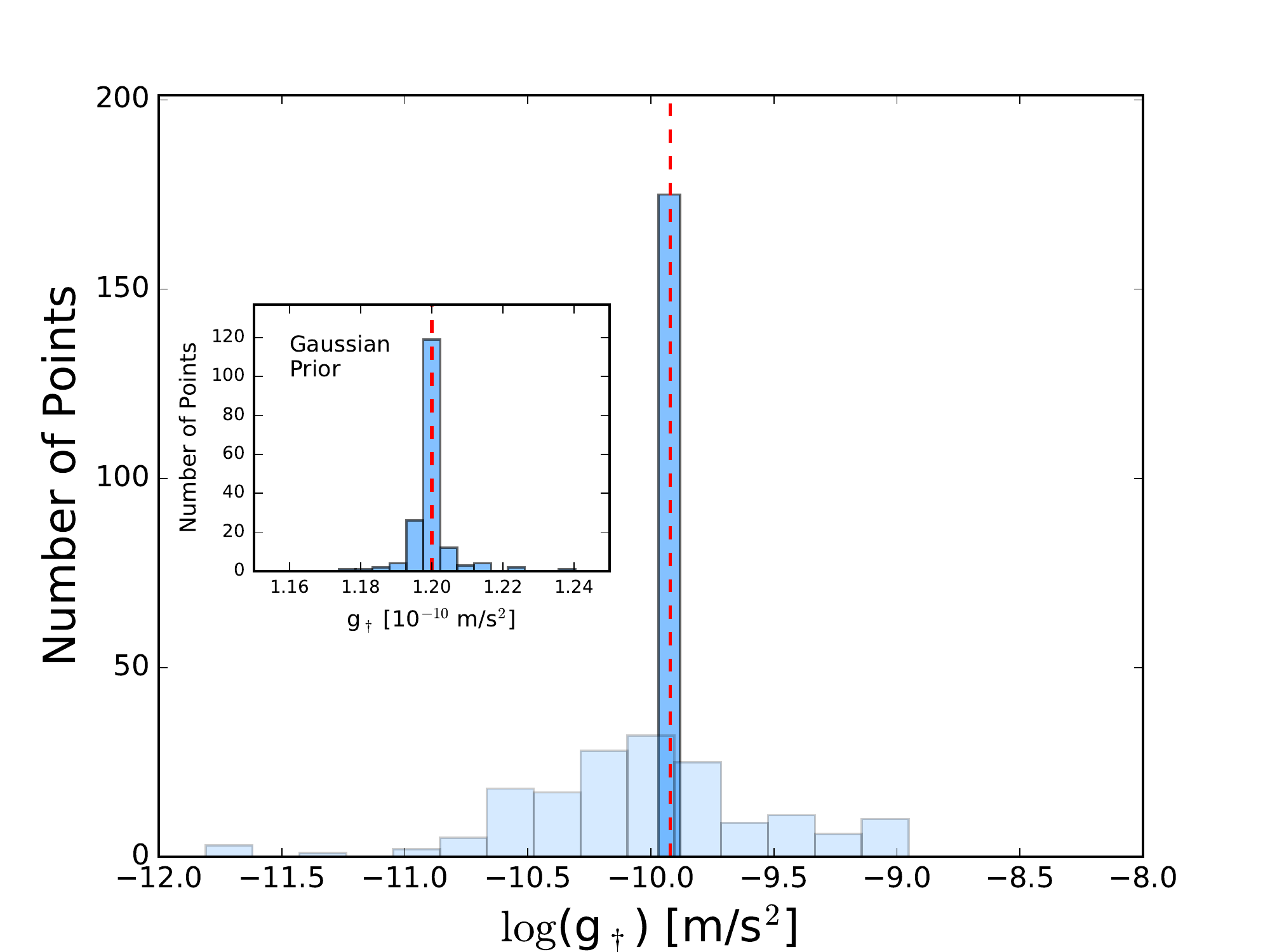}
\caption{Distribution of optimal $g_\dagger$ imposing a flat (light blue) and Gaussian (dark blue) prior. The red dashed line marks our fiducial value 1.2 $\times$ 10$^{-10}$ m/s$^2$. The inset shows the Gaussian prior alone, zoomed in and switched to a linear scale to resolve the distribution. Note the vast difference in scales: the flat prior results in a broad range of $g_\dagger$ (although with no improvement in $\chi^2$ , as seen in Figure \ref{CDF_Chi}), while the Gaussian prior effectively returns a constant $g_\dagger$.}
\label{gdagger}
\end{figure}

We show the distributions of best-fit g$_\dagger$ in Figure \ref{gdagger} for both flat and Gaussian priors. The flat prior leads to a wide distribution of g$_\dagger$ , while a Gaussian prior results in a tight distribution around its fiducial value. The Gaussian prior indeed results in a distribution so close to a fixed g$_\dagger$, with a width smaller than the standard deviation imposed by the prior, that it appears as a $\delta$-function on a scale that accommodates the wide distribution of the flat prior. That the apparent distribution of g$_\dagger$ is so large in the case of the flat prior is indicative of parameter degeneracy: changes in g$_\dagger$ can be compensated for by changes in the mass-to-light ratio (or the nuisance parameters) so that both may vary in an unphysical way to achieve trivial gains in $\chi^2$.

Adjusting the value of $g_{\dagger}$ improves neither the fits nor the rms scatter.
The data are consistent with the same value for all galaxies.
There is no need to invoke variable $g_{\dagger}$ \citep[cf.\ ][]{Bottema2015}.
To do so would violate the law of parsimony (Occam's razor).

\section{Discussion and conclusion}

By fitting individual galaxies with the MCMC method, we showed that the intrinsic scatter of the RAR is extremely small. The baryonic matter distribution can reproduce the rotation curve very well, and vice versa.

The tightness of the RAR provides a challenge for the standard $\Lambda$CDM cosmology. Recent studies claim that the RAR is a natural product of galaxy formation in $\Lambda$CDM \citep{2017ApJ...835L..17K, 2017PhRvL.118p1103L}, but none of these studies have properly taken into account observational effects when comparing theory and observations. There are two major issues: (1) general confusion between the concepts of observed and intrinsic scatter, and (2) oversampling of simulated rotation curves. \citet{2017ApJ...835L..17K} analyzed 32 galaxies from the MUGS2 ``zoom-in'' hydrodynamic simulations and argued that the dissipative collapse of baryons can result in a relation with a scatter of 0.06 dex. Similarly, \citet{2017PhRvL.118p1103L} analyzed a suite of simulated galaxies from the EAGLE project and found a relation with a scatter of 0.09 dex. However, these values {cannot} be compared with the observed scatter because (1) measurement errors are {not} added to the simulated galaxies and are not properly propagated, and (2) the simulated rotation curves are {not} sampled with the same number of resolution elements as in the observations. Both effects can significantly underpredict the scatter expected from cosmological simulations. Oversampling the same error-free simulated galaxy over and over can artificially decrease the expected scatter around the mean relation.

In contrast, \citet{Desmond2017} took both radial sampling and observational errors into account (see also \citealt{2016MNRAS.456L.127D}). \citet{Desmond2017} found that fiducial $\Lambda$CDM models significantly overpredict the observed scatter around the mass discrepancy-acceleration relation by $\sim$3.5 $\sigma$. This discrepancy remains even if one assumes a perfect 1:1 relation between halo mass and stellar mass of galaxies with no scatter. Hence, this problem seems to be due to the stochastic hierarchical formation of DM halos, independent of baryonic physics.

The detailed shape of the RAR is also important. \citet{2017PhRvL.118p1103L} fit the Equation \ref{mond} to their simulated galaxies, but found a value of $g_{\dag}=2.6 \times 10^{-10}$ m s$^{-2}$ instead of $g_{\dag}=1.2 \pm 0.02 \times 10^{-10}$ m s$^{-2}$. This is a 70 $\sigma$ discrepancy. Recently, \citet{2017arXiv170305287T} analyzed galaxies from the MassiveBlack-II simulation. They also found a correlation between $g_{\rm tot}$ and $g_{\rm bar}$, but this is better fit by a power law with a width of $\sim$0.1 dex rather than by Equation 3. Hence, the detailed properties of the RAR (shape and scatter) remain an open issue for $\Lambda$CDM models of galaxy formation.

We here reported an rms scatter of 0.057 dex after marginalizing over the uncertainties due to mass-to-light ratio, galaxy distance, and disk inclination. This observed scatter is a hard upper limit to the intrinsic scatter since the errors on the observed velocities are non-negligible. This is hard to understand in a $\Lambda$CDM scenario since the diverse formation histories of galaxies must necessarily introduce some scatter on the relation, as demonstrated by the existing cosmological simulations.

In order to reconcile the conundrum that the standard cold dark matter faces, some new dark sectors have been proposed, such
as dipolar DM particles subjected to gravitational polarization \citep{2009PhRvD..80b3524B, 2008PhRvD..78b4031B}, dark fluids \citep{2010ApJ...712..130Z, 2015PhRvD..91b4022K}, dissipative DM particles \citep{2017PhRvD..95b3009C}, or fifth forces \citep{2017PhRvD..95f4050B}. To be viable, such hypotheses must explain the shape, amplitude, and negligible scatter of the RAR. The coupling between the baryonic matter and dark matter must be rather strong to explain these observations.

On the other hand, the tight RAR could be easily understood in MOND \citep{1983ApJ...270..371M}. MOND dictates that the equations of motions become scale-invariant at accelerations $a < a_0$, where $a_0$ corresponds to $\mathrm{g_\dagger}$ in Equation \ref{mond} \citep{2009ApJ...698.1630M}. Thus, Equation \ref{mond} is related to the interpolation function $\nu(g_\mathrm{bar}/a_0)$ of MOND. The scale invariance can be achieved in two ways: modified gravity (MG) by changing the Poisson equation \citep{1984ApJ...286....7B, 2010MNRAS.403..886M}, and modified inertia (MI) by changing the Second Law of Newton \citep{1994AnPhy.229..384M}. MI requires the relation $g_\mathrm{obs} = \nu(g_\mathrm{bar}/a_0) g_\mathrm{bar}$ to be true for circular orbits, leading to zero intrinsic scatter in the RAR of rotating disk galaxies \citep{1994AnPhy.229..384M}. MG requires the system to be spherically symmetric to hold precisely to the acceleration relation \citep{1984ApJ...286....7B}. Thus, the predicted $g_{\rm tot}$ of disk galaxies can show subtle differences in MG \citep{1995MNRAS.276..453B,Milgrom2012}, and some non-zero intrinsic scatter could be introduced. Our results suggest a preference for the MI theory.

Other modified gravity theories such as emergent gravity could also potentially explain the RAR \citep{2016arXiv161102269V}. However, \citet{2017MNRAS.468L..68L} shows that the RAR predicted by this theory has significant intrinsic scatter and the residuals should correlate with radius, which contradicts the data.

The extremely small intrinsic scatter of the RAR provides a tool for testing various theories of modified gravity or dark matter. It also provides key insights toward the path for finally solving the ``dark matter problem''.

\begin{acknowledgements}
We thank Harley Katz for useful discussion about using MCMC and GetDist. This work is based in part on observations made with the Spitzer Space Telescope, which is operated by the Jet Propulsion Laboratory, 
California Institute of Technology under a contract with NASA. This publication was made possible in part through the support of the John Templeton Foundation. 
\end{acknowledgements}

%\begin{appendix}
%\section{Additional material}
\longtab[1]{
\begin{longtable}{cccccccccc}
\caption{Maximum posterior parameters and reduced $\chi^2$ of individual rotation curve fits to the RAR. L$_{[3.6]}$, $D_0$ and $i_0$ are the original luminosity, distance, and inclination from the SPARC database. Galaxies are ordered by decreasing luminosity.}
\label{tab:galaxy}\\
\hline\hline
{SPARC ID} & {Galaxy name} & {$\log$(L$_{[3.6]}$)} & {$\mathrm{\Upsilon_{disk}}$} & {$\mathrm{\Upsilon_{bulge}}$} & {Distance} & {D/D$_0$} & {Inclination} & {$i/i_0$} & {$\chi^2_\nu$}\\
{} &
{} &
{(L$_\odot$)}&
{(M$_\sun$/L$_\sun$)} &
{(M$_\sun$/L$_\sun$)} &
{(Mpc)} &
{} &
{(deg.)} &
{} &
{}\\
\hline
\endfirsthead
\caption{Continued.}\\
\hline
{SPARC ID} & {Galaxy name}  & {$\log$(L$_{[3.6]}$)} & {$\mathrm{\Upsilon_{disk}}$} & {$\mathrm{\Upsilon_{bulge}}$} & {Distance} & {D/D$_0$} & {Inclination} & {$i/i_0$} & {$\chi^2_\nu$}\\
{} &
{} &
{(L$_\odot$)}&
{(M$_\sun$/L$_\sun$)} &
{(M$_\sun$/L$_\sun$)} &
{(Mpc)} &
{} &
{(deg.)} &
{} &
{}\\
\hline
\endhead
\hline
\endfoot
\hline
\endlastfoot
001 & UGC02487 & 11.69 & 1.83 $\pm$ 0.20 & 0.91 $\pm$ 0.18 & 63.7 $\pm$ 9.4 & 0.92 & 31.2 $\pm$ 2.8 & 0.87 & 4.482 \\ 
002 & UGC02885 & 11.61 & 0.45 $\pm$ 0.06 & 0.97 $\pm$ 0.08 & 82.7 $\pm$ 5.5 & 1.03 & 64.7 $\pm$ 3.4 & 1.01 & 0.858 \\ 
003 & NGC6195 & 11.59 & 0.32 $\pm$ 0.06 & 0.85 $\pm$ 0.07 & 115 $\pm$ 11 & 0.90 & 57.0 $\pm$ 4.2 & 0.92 & 2.258 \\ 
004 & UGC11455 & 11.57 & 0.38 $\pm$ 0.04 & \dots & 84.7 $\pm$ 5.4 & 1.08 & 90.0 $\pm$ 0.6 & 1.00 & 6.545 \\ 
005 & NGC5371 & 11.53 & 3.30 $\pm$ 0.29 & \dots & 7.44 $\pm$ 0.68 & 0.19 & 52.7 $\pm$ 2.0 & 0.99 & 10.156 \\ 
006 & NGC2955 & 11.50 & 0.37 $\pm$ 0.06 & 0.84 $\pm$ 0.08 & 95.3 $\pm$ 8.9 & 0.97 & 52.7 $\pm$ 4.3 & 0.94 & 3.906 \\ 
007 & NGC0801 & 11.49 & 1.33 $\pm$ 0.12 & \dots & 33.0 $\pm$ 2.0 & 0.41 & 79.9 $\pm$ 1.0 & 1.00 & 7.753 \\ 
008 & ESO563-G021 & 11.49 & 0.43 $\pm$ 0.04 & \dots & 88.0 $\pm$ 4.9 & 1.45 & 84.0 $\pm$ 2.7 & 1.01 & 28.836 \\ 
009 & UGC09133 & 11.45 & 1.64 $\pm$ 0.06 & 1.10 $\pm$ 0.02 & 36.3 $\pm$ 6.9 & 0.64 & 45.7 $\pm$ 5.4 & 0.86 & 6.937 \\ 
010 & UGC02953 & 11.41 & 0.61 $\pm$ 0.01 & 0.62 $\pm$ 0.01 & 18.1 $\pm$ 1.9 & 1.10 & 50.4 $\pm$ 3.5 & 1.01 & 5.661 \\ 
011 & NGC7331 & 11.40 & 0.32 $\pm$ 0.02 & 0.60 $\pm$ 0.12 & 15.7 $\pm$ 0.6 & 1.07 & 75.3 $\pm$ 2.0 & 1.00 & 1.289 \\ 
012 & NGC3992 & 11.36 & 0.76 $\pm$ 0.10 & \dots & 21.3 $\pm$ 1.7 & 0.90 & 55.1 $\pm$ 1.9 & 0.98 & 3.465 \\ 
013 & NGC6674 & 11.33 & 0.95 $\pm$ 0.11 & 1.30 $\pm$ 0.45 & 43.2 $\pm$ 6.8 & 0.84 & 50.5 $\pm$ 5.2 & 0.94 & 10.638 \\ 
014 & NGC5985 & 11.32 & 0.63 $\pm$ 0.10 & 3.32 $\pm$ 0.30 & 46.7 $\pm$ 4.1 & 1.18 & 60.3 $\pm$ 2.0 & 1.01 & 6.974 \\ 
015 & NGC2841 & 11.27 & 0.81 $\pm$ 0.05 & 0.93 $\pm$ 0.05 & 15.5 $\pm$ 0.6 & 1.10 & 81.9 $\pm$ 5.2 & 1.08 & 1.515 \\ 
016 & IC4202 & 11.25 & 1.60 $\pm$ 0.19 & 0.34 $\pm$ 0.04 & 64.4 $\pm$ 5.2 & 0.64 & 90.0 $\pm$ 0.6 & 1.00 & 41.908 \\ 
017 & NGC5005 & 11.25 & 0.54 $\pm$ 0.07 & 0.56 $\pm$ 0.07 & 16.6 $\pm$ 1.3 & 0.98 & 67.9 $\pm$ 2.0 & 1.00 & 0.091 \\ 
018 & NGC5907 & 11.24 & 1.08 $\pm$ 0.07 & \dots & 10.5 $\pm$ 0.4 & 0.61 & 87.5 $\pm$ 1.8 & 0.99 & 7.730 \\ 
019 & UGC05253 & 11.23 & 0.63 $\pm$ 0.05 & 0.69 $\pm$ 0.03 & 22.5 $\pm$ 3.5 & 0.98 & 36.8 $\pm$ 3.2 & 1.00 & 4.747 \\ 
020 & NGC5055 & 11.18 & 0.56 $\pm$ 0.01 & \dots & 9.83 $\pm$ 0.30 & 0.99 & 43.8 $\pm$ 0.9 & 0.80 & 7.415 \\ 
021 & NGC2998 & 11.18 & 0.82 $\pm$ 0.10 & \dots & 48.8 $\pm$ 3.8 & 0.72 & 57.2 $\pm$ 2.0 & 0.99 & 2.940 \\ 
022 & UGC11914 & 11.18 & 0.22 $\pm$ 0.03 & 0.48 $\pm$ 0.04 & 24.0 $\pm$ 3.4 & 1.42 & 38.4 $\pm$ 3.1 & 1.24 & 1.731 \\ 
023 & NGC3953 & 11.15 & 0.59 $\pm$ 0.10 & \dots & 16.0 $\pm$ 1.8 & 0.89 & 61.9 $\pm$ 1.0 & 1.00 & 3.424 \\ 
024 & UGC12506 & 11.14 & 1.12 $\pm$ 0.16 & \dots & 85.3 $\pm$ 6.3 & 0.85 & 85.4 $\pm$ 3.4 & 0.99 & 1.981 \\ 
025 & NGC0891 & 11.14 & 0.33 $\pm$ 0.02 & 0.40 $\pm$ 0.05 & 11.4 $\pm$ 0.4 & 1.15 & 90.0 $\pm$ 0.6 & 1.00 & 7.368 \\ 
026 & UGC06614 & 11.09 & 0.51 $\pm$ 0.12 & 0.50 $\pm$ 0.09 & 88.6 $\pm$ 8.8 & 1.00 & 32.8 $\pm$ 2.6 & 0.91 & 1.164 \\ 
027 & UGC02916 & 11.09 & 1.57 $\pm$ 0.24 & 0.73 $\pm$ 0.06 & 54.1 $\pm$ 7.9 & 0.83 & 44.5 $\pm$ 4.1 & 0.89 & 11.652 \\ 
028 & UGC03205 & 11.06 & 0.73 $\pm$ 0.06 & 1.32 $\pm$ 0.07 & 42.5 $\pm$ 3.1 & 0.85 & 66.2 $\pm$ 3.9 & 0.99 & 4.196 \\ 
029 & NGC5033 & 11.04 & 1.03 $\pm$ 0.08 & 0.43 $\pm$ 0.06 & 12.7 $\pm$ 0.5 & 0.81 & 65.9 $\pm$ 1.0 & 1.00 & 8.024 \\ 
030 & NGC4088 & 11.03 & 0.40 $\pm$ 0.07 & \dots & 13.4 $\pm$ 1.3 & 0.74 & 68.4 $\pm$ 2.0 & 0.99 & 0.664 \\ 
031 & NGC4157 & 11.02 & 0.43 $\pm$ 0.06 & 0.64 $\pm$ 0.15 & 15.7 $\pm$ 1.3 & 0.87 & 81.7 $\pm$ 3.0 & 1.00 & 0.720 \\ 
032 & UGC03546 & 11.01 & 0.68 $\pm$ 0.08 & 0.51 $\pm$ 0.04 & 25.4 $\pm$ 3.4 & 0.88 & 54.1 $\pm$ 4.5 & 0.98 & 0.907 \\ 
033 & UGC06787 & 10.99 & 0.45 $\pm$ 0.04 & 0.28 $\pm$ 0.01 & 37.7 $\pm$ 1.8 & 1.77 & 68.6 $\pm$ 2.7 & 1.04 & 20.814 \\ 
034 & NGC4051 & 10.98 & 0.45 $\pm$ 0.09 & \dots & 15.3 $\pm$ 1.9 & 0.85 & 47.1 $\pm$ 2.8 & 0.96 & 2.491 \\ 
035 & NGC4217 & 10.93 & 1.17 $\pm$ 0.20 & 0.17 $\pm$ 0.02 & 18.2 $\pm$ 1.5 & 1.01 & 86.1 $\pm$ 1.9 & 1.00 & 3.171 \\ 
036 & NGC3521 & 10.93 & 0.46 $\pm$ 0.05 & \dots & 8.54 $\pm$ 0.83 & 1.11 & 75.3 $\pm$ 4.9 & 1.00 & 0.510 \\ 
037 & NGC2903 & 10.91 & 0.21 $\pm$ 0.01 & \dots & 11.0 $\pm$ 0.5 & 1.67 & 67.6 $\pm$ 2.8 & 1.02 & 20.637 \\ 
038 & NGC2683 & 10.91 & 0.55 $\pm$ 0.06 & 0.73 $\pm$ 0.18 & 9.44 $\pm$ 0.46 & 0.96 & 77.0 $\pm$ 5.2 & 0.96 & 3.370 \\ 
039 & NGC4013 & 10.90 & 0.35 $\pm$ 0.05 & 0.79 $\pm$ 0.17 & 18.1 $\pm$ 1.0 & 1.01 & 89.0 $\pm$ 0.8 & 1.00 & 1.807 \\ 
040 & NGC7814 & 10.87 & 1.17 $\pm$ 0.12 & 0.52 $\pm$ 0.04 & 15.8 $\pm$ 0.6 & 1.10 & 90.0 $\pm$ 0.6 & 1.00 & 1.334 \\ 
041 & UGC06786 & 10.87 & 0.27 $\pm$ 0.02 & 0.34 $\pm$ 0.01 & 57.2 $\pm$ 2.6 & 1.95 & 67.7 $\pm$ 2.6 & 1.06 & 1.389 \\ 
042 & NGC3877 & 10.86 & 0.40 $\pm$ 0.07 & \dots & 16.8 $\pm$ 1.7 & 0.93 & 76.0 $\pm$ 1.0 & 1.00 & 10.221 \\ 
043 & NGC0289 & 10.86 & 0.92 $\pm$ 0.09 & \dots & 14.8 $\pm$ 2.9 & 0.71 & 42.9 $\pm$ 4.7 & 0.93 & 2.132 \\ 
044 & NGC1090 & 10.86 & 0.74 $\pm$ 0.07 & \dots & 24.3 $\pm$ 1.8 & 0.66 & 63.5 $\pm$ 3.0 & 0.99 & 2.778 \\ 
045 & NGC3726 & 10.85 & 0.47 $\pm$ 0.07 & \dots & 13.8 $\pm$ 1.3 & 0.77 & 52.0 $\pm$ 2.0 & 0.98 & 2.982 \\ 
046 & UGC09037 & 10.84 & 0.20 $\pm$ 0.02 & \dots & 80.2 $\pm$ 5.9 & 0.96 & 63.3 $\pm$ 4.1 & 0.97 & 2.259 \\ 
047 & NGC6946 & 10.82 & 0.64 $\pm$ 0.05 & 0.71 $\pm$ 0.04 & 4.18 $\pm$ 0.44 & 0.76 & 37.7 $\pm$ 2.0 & 0.99 & 1.525 \\ 
048 & NGC4100 & 10.77 & 0.76 $\pm$ 0.10 & \dots & 15.0 $\pm$ 1.2 & 0.83 & 72.7 $\pm$ 2.0 & 1.00 & 1.658 \\ 
049 & NGC3893 & 10.77 & 0.45 $\pm$ 0.06 & \dots & 19.4 $\pm$ 1.7 & 1.08 & 49.5 $\pm$ 1.9 & 1.01 & 0.997 \\ 
050 & UGC06973 & 10.73 & 0.17 $\pm$ 0.02 & 0.39 $\pm$ 0.07 & 23.0 $\pm$ 1.6 & 1.28 & 72.7 $\pm$ 2.8 & 1.02 & 15.579 \\ 
051 & ESO079-G014 & 10.71 & 0.50 $\pm$ 0.09 & \dots & 31.6 $\pm$ 3.0 & 1.10 & 79.4 $\pm$ 4.7 & 1.00 & 4.334 \\ 
052 & UGC08699 & 10.70 & 0.63 $\pm$ 0.10 & 0.70 $\pm$ 0.05 & 40.4 $\pm$ 4.4 & 1.03 & 73.2 $\pm$ 7.5 & 1.00 & 0.989 \\ 
053 & NGC4138 & 10.64 & 0.55 $\pm$ 0.11 & 0.69 $\pm$ 0.17 & 17.9 $\pm$ 1.8 & 0.99 & 53.0 $\pm$ 2.8 & 1.00 & 2.492 \\ 
054 & NGC3198 & 10.58 & 0.77 $\pm$ 0.03 & \dots & 10.4 $\pm$ 0.4 & 0.75 & 71.1 $\pm$ 3.1 & 0.97 & 2.057 \\ 
055 & NGC3949 & 10.58 & 0.44 $\pm$ 0.07 & \dots & 17.3 $\pm$ 1.9 & 0.96 & 54.8 $\pm$ 2.0 & 1.00 & 0.547 \\ 
056 & NGC6015 & 10.51 & 1.12 $\pm$ 0.06 & \dots & 11.8 $\pm$ 0.6 & 0.69 & 59.8 $\pm$ 2.0 & 1.00 & 10.873 \\ 
057 & NGC3917 & 10.34 & 0.55 $\pm$ 0.09 & \dots & 16.9 $\pm$ 1.5 & 0.94 & 78.9 $\pm$ 2.0 & 1.00 & 4.603 \\ 
058 & NGC4085 & 10.34 & 0.35 $\pm$ 0.06 & \dots & 16.8 $\pm$ 1.8 & 0.93 & 81.9 $\pm$ 2.0 & 1.00 & 9.088 \\ 
\clearpage
059 & NGC4389 & 10.33 & 0.30 $\pm$ 0.07 & \dots & 11.6 $\pm$ 2.2 & 0.64 & 42.8 $\pm$ 4.1 & 0.86 & 9.313 \\ 
060 & NGC4559 & 10.29 & 0.52 $\pm$ 0.06 & \dots & 6.66 $\pm$ 0.36 & 0.74 & 67.0 $\pm$ 1.0 & 1.00 & 0.496 \\ 
061 & NGC3769 & 10.27 & 0.41 $\pm$ 0.07 & \dots & 16.2 $\pm$ 1.3 & 0.90 & 69.8 $\pm$ 2.0 & 1.00 & 0.949 \\ 
062 & NGC4010 & 10.24 & 0.36 $\pm$ 0.07 & \dots & 18.2 $\pm$ 1.5 & 1.01 & 89.0 $\pm$ 0.8 & 1.00 & 2.741 \\ 
063 & NGC3972 & 10.16 & 0.50 $\pm$ 0.08 & \dots & 19.2 $\pm$ 1.7 & 1.07 & 77.0 $\pm$ 1.0 & 1.00 & 2.074 \\ 
064 & UGC03580 & 10.12 & 0.29 $\pm$ 0.02 & 0.11 $\pm$ 0.01 & 22.4 $\pm$ 1.6 & 1.08 & 63.4 $\pm$ 3.8 & 1.01 & 2.291 \\ 
065 & NGC6503 & 10.11 & 0.45 $\pm$ 0.02 & \dots & 6.47 $\pm$ 0.16 & 1.03 & 74.6 $\pm$ 1.8 & 1.01 & 2.979 \\ 
066 & UGC11557 & 10.08 & 0.42 $\pm$ 0.11 & \dots & 22.5 $\pm$ 6.2 & 0.93 & 19.6 $\pm$ 4.0 & 0.65 & 3.175 \\ 
067 & UGC00128 & 10.08 & 2.49 $\pm$ 0.12 & \dots & 58.1 $\pm$ 9.3 & 0.90 & 40.7 $\pm$ 4.3 & 0.71 & 6.254 \\ 
068 & F579-V1 & 10.07 & 0.63 $\pm$ 0.14 & \dots & 89.6 $\pm$ 8.9 & 1.00 & 26.3 $\pm$ 2.2 & 1.01 & 2.559 \\ 
069 & NGC4183 & 10.03 & 0.79 $\pm$ 0.14 & \dots & 13.0 $\pm$ 0.9 & 0.72 & 81.8 $\pm$ 2.0 & 1.00 & 1.132 \\ 
070 & F571-8 & 10.01 & 0.11 $\pm$ 0.02 & \dots & 102 $\pm$ 8 & 1.92 & 87.1 $\pm$ 2.9 & 1.02 & 41.610 \\ 
071 & NGC2403 & 10.00 & 0.51 $\pm$ 0.01 & \dots & 3.33 $\pm$ 0.11 & 1.05 & 66.1 $\pm$ 2.1 & 1.05 & 14.142 \\ 
072 & UGC06930 & 9.95 & 0.63 $\pm$ 0.13 & \dots & 17.3 $\pm$ 2.3 & 0.96 & 28.8 $\pm$ 2.5 & 0.90 & 1.233 \\ 
073 & F568-3 & 9.92 & 0.41 $\pm$ 0.09 & \dots & 82.1 $\pm$ 8.1 & 1.00 & 38.3 $\pm$ 3.2 & 0.96 & 3.064 \\ 
074 & UGC01230 & 9.88 & 0.72 $\pm$ 0.17 & \dots & 53.1 $\pm$ 10.6 & 0.99 & 18.4 $\pm$ 2.4 & 0.84 & 2.951 \\ 
075 & NGC0247 & 9.87 & 0.78 $\pm$ 0.08 & \dots & 3.23 $\pm$ 0.16 & 0.87 & 68.9 $\pm$ 3.1 & 0.93 & 3.060 \\ 
076 & NGC7793 & 9.85 & 0.55 $\pm$ 0.09 & \dots & 3.60 $\pm$ 0.18 & 1.00 & 43.3 $\pm$ 2.9 & 0.92 & 1.013 \\ 
077 & UGC06917 & 9.83 & 0.54 $\pm$ 0.09 & \dots & 18.4 $\pm$ 1.5 & 1.02 & 56.1 $\pm$ 1.9 & 1.00 & 1.315 \\ 
078 & NGC1003 & 9.83 & 0.37 $\pm$ 0.03 & \dots & 10.8 $\pm$ 0.9 & 0.94 & 66.9 $\pm$ 4.9 & 1.00 & 4.669 \\ 
079 & F574-1 & 9.82 & 0.71 $\pm$ 0.13 & \dots & 95.7 $\pm$ 8.7 & 0.99 & 62.6 $\pm$ 6.6 & 0.96 & 2.501 \\ 
080 & F568-1 & 9.80 & 0.61 $\pm$ 0.13 & \dots & 95.2 $\pm$ 8.8 & 1.05 & 32.4 $\pm$ 2.2 & 1.24 & 1.287 \\ 
081 & UGC06983 & 9.72 & 0.77 $\pm$ 0.11 & \dots & 17.3 $\pm$ 1.2 & 0.96 & 49.0 $\pm$ 1.0 & 1.00 & 1.392 \\ 
082 & UGC05986 & 9.67 & 0.31 $\pm$ 0.04 & \dots & 14.6 $\pm$ 0.8 & 1.69 & 90.0 $\pm$ 1.8 & 1.00 & 3.997 \\ 
083 & NGC0055 & 9.67 & 0.19 $\pm$ 0.03 & \dots & 1.74 $\pm$ 0.07 & 0.82 & 71.3 $\pm$ 3.2 & 0.93 & 1.579 \\ 
084 & ESO116-G012 & 9.63 & 0.35 $\pm$ 0.04 & \dots & 17.9 $\pm$ 0.9 & 1.38 & 74.5 $\pm$ 2.9 & 1.01 & 2.444 \\ 
085 & UGC07323 & 9.61 & 0.41 $\pm$ 0.09 & \dots & 6.87 $\pm$ 0.96 & 0.86 & 46.5 $\pm$ 3.0 & 0.99 & 0.660 \\ 
086 & UGC05005 & 9.61 & 0.45 $\pm$ 0.10 & \dots & 50.8 $\pm$ 10.1 & 0.95 & 34.6 $\pm$ 4.8 & 0.84 & 0.315 \\ 
087 & F561-1 & 9.61 & 0.52 $\pm$ 0.13 & \dots & 65.2 $\pm$ 10.1 & 0.98 & 13.5 $\pm$ 1.4 & 0.56 & 1.564 \\ 
088 & NGC0024 & 9.59 & 1.01 $\pm$ 0.11 & \dots & 7.55 $\pm$ 0.32 & 1.03 & 66.1 $\pm$ 2.6 & 1.03 & 0.850 \\ 
089 & F568-V1 & 9.58 & 0.81 $\pm$ 0.16 & \dots & 83.7 $\pm$ 7.4 & 1.04 & 51.1 $\pm$ 4.4 & 1.28 & 1.042 \\ 
090 & UGC06628 & 9.57 & 0.52 $\pm$ 0.13 & \dots & 14.4 $\pm$ 4.7 & 0.95 & 10.6 $\pm$ 2.9 & 0.53 & 0.851 \\ 
091 & UGC02455 & 9.56 & 0.33 $\pm$ 0.09 & \dots & 2.01 $\pm$ 0.50 & 0.29 & 49.3 $\pm$ 5.3 & 0.97 & 6.549 \\ 
092 & UGC07089 & 9.55 & 0.36 $\pm$ 0.08 & \dots & 13.3 $\pm$ 1.2 & 0.74 & 79.4 $\pm$ 3.1 & 0.99 & 0.426 \\ 
093 & UGC05999 & 9.53 & 0.48 $\pm$ 0.11 & \dots & 47.2 $\pm$ 9.3 & 0.99 & 19.6 $\pm$ 2.4 & 0.89 & 5.693 \\ 
094 & NGC2976 & 9.53 & 0.35 $\pm$ 0.08 & \dots & 3.58 $\pm$ 0.18 & 1.00 & 62.5 $\pm$ 6.4 & 1.02 & 1.730 \\ 
095 & UGC05750 & 9.52 & 0.48 $\pm$ 0.11 & \dots & 47.5 $\pm$ 9.6 & 0.81 & 53.9 $\pm$ 8.7 & 0.84 & 1.352 \\ 
096 & NGC0100 & 9.51 & 0.28 $\pm$ 0.06 & \dots & 15.9 $\pm$ 1.5 & 1.18 & 89.0 $\pm$ 0.8 & 1.00 & 1.286 \\ 
097 & UGC00634 & 9.48 & 0.49 $\pm$ 0.09 & \dots & 31.3 $\pm$ 6.2 & 1.01 & 37.4 $\pm$ 4.7 & 1.01 & 2.425 \\ 
098 & F563-V2 & 9.48 & 0.59 $\pm$ 0.14 & \dots & 63.4 $\pm$ 10.5 & 1.06 & 36.8 $\pm$ 4.3 & 1.27 & 0.991 \\ 
099 & NGC5585 & 9.47 & 0.22 $\pm$ 0.01 & \dots & 7.81 $\pm$ 0.46 & 1.11 & 51.2 $\pm$ 1.9 & 1.00 & 6.817 \\ 
100 & NGC0300 & 9.47 & 0.40 $\pm$ 0.05 & \dots & 2.09 $\pm$ 0.10 & 1.00 & 47.4 $\pm$ 2.1 & 1.13 & 0.906 \\ 
101 & UGC06923 & 9.46 & 0.42 $\pm$ 0.09 & \dots & 16.5 $\pm$ 1.5 & 0.92 & 64.7 $\pm$ 2.0 & 1.00 & 1.624 \\ 
102 & F574-2 & 9.46 & 0.49 $\pm$ 0.12 & \dots & 88.1 $\pm$ 9.0 & 0.99 & 13.1 $\pm$ 1.8 & 0.44 & 0.092 \\ 
103 & UGC07125 & 9.43 & 0.92 $\pm$ 0.15 & \dots & 7.45 $\pm$ 0.40 & 0.38 & 89.9 $\pm$ 1.8 & 1.00 & 1.599 \\ 
104 & UGC07524 & 9.39 & 0.79 $\pm$ 0.12 & \dots & 4.50 $\pm$ 0.23 & 0.95 & 38.8 $\pm$ 1.6 & 0.84 & 1.839 \\ 
105 & UGC06399 & 9.36 & 0.53 $\pm$ 0.10 & \dots & 19.0 $\pm$ 1.5 & 1.05 & 75.1 $\pm$ 2.0 & 1.00 & 0.520 \\ 
106 & UGC07151 & 9.36 & 0.50 $\pm$ 0.05 & \dots & 6.35 $\pm$ 0.28 & 0.92 & 90.0 $\pm$ 2.0 & 1.00 & 3.751 \\ 
107 & F567-2 & 9.33 & 0.56 $\pm$ 0.13 & \dots & 78.2 $\pm$ 11.8 & 0.99 & 14.5 $\pm$ 1.7 & 0.73 & 2.204 \\ 
108 & UGC04325 & 9.31 & 0.94 $\pm$ 0.19 & \dots & 10.2 $\pm$ 1.4 & 1.07 & 41.3 $\pm$ 2.7 & 1.01 & 9.429 \\ 
109 & UGC00191 & 9.30 & 1.10 $\pm$ 0.13 & \dots & 12.5 $\pm$ 2.5 & 0.73 & 43.0 $\pm$ 4.7 & 0.96 & 3.842 \\ 
110 & F563-1 & 9.28 & 0.56 $\pm$ 0.12 & \dots & 51.7 $\pm$ 8.2 & 1.06 & 27.4 $\pm$ 2.5 & 1.10 & 1.499 \\ 
111 & F571-V1 & 9.27 & 0.50 $\pm$ 0.12 & \dots & 80.0 $\pm$ 8.0 & 1.00 & 27.8 $\pm$ 2.0 & 0.93 & 0.288 \\ 
112 & UGC07261 & 9.24 & 0.56 $\pm$ 0.12 & \dots & 12.8 $\pm$ 3.4 & 0.98 & 29.0 $\pm$ 5.0 & 0.97 & 0.827 \\ 
113 & UGC10310 & 9.24 & 0.62 $\pm$ 0.14 & \dots & 13.4 $\pm$ 3.3 & 0.88 & 31.7 $\pm$ 4.4 & 0.93 & 1.762 \\ 
114 & UGC02259 & 9.24 & 1.14 $\pm$ 0.19 & \dots & 10.0 $\pm$ 1.3 & 0.95 & 40.9 $\pm$ 2.8 & 1.00 & 7.221 \\ 
115 & F583-4 & 9.23 & 0.48 $\pm$ 0.11 & \dots & 50.3 $\pm$ 8.8 & 0.94 & 51.2 $\pm$ 7.0 & 0.93 & 0.134 \\ 
116 & UGC12732 & 9.22 & 1.07 $\pm$ 0.14 & \dots & 11.4 $\pm$ 2.6 & 0.86 & 36.8 $\pm$ 4.7 & 0.94 & 0.496 \\ 
117 & UGC06818 & 9.20 & 0.29 $\pm$ 0.06 & \dots & 14.8 $\pm$ 1.6 & 0.82 & 74.3 $\pm$ 3.1 & 0.99 & 5.387 \\ 
118 & UGC04499 & 9.19 & 0.51 $\pm$ 0.10 & \dots & 9.85 $\pm$ 1.06 & 0.79 & 49.5 $\pm$ 3.0 & 0.99 & 1.776 \\ 
\clearpage
119 & F563-V1 & 9.19 & 0.48 $\pm$ 0.12 & \dots & 39.3 $\pm$ 13.6 & 0.73 & 29.4 $\pm$ 10.0 & 0.49 & 0.875 \\ 
120 & UGC06667 & 9.15 & 1.00 $\pm$ 0.20 & \dots & 23.8 $\pm$ 1.3 & 1.32 & 89.0 $\pm$ 0.8 & 1.00 & 5.357 \\ 
121 & UGC02023 & 9.12 & 0.49 $\pm$ 0.12 & \dots & 10.0 $\pm$ 3.1 & 0.96 & 14.3 $\pm$ 3.4 & 0.75 & 1.147 \\ 
122 & UGC04278 & 9.12 & 0.53 $\pm$ 0.07 & \dots & 10.3 $\pm$ 0.7 & 1.08 & 89.9 $\pm$ 1.8 & 1.00 & 2.597 \\ 
123 & UGC12632 & 9.11 & 1.08 $\pm$ 0.19 & \dots & 6.72 $\pm$ 0.84 & 0.69 & 45.2 $\pm$ 3.0 & 0.98 & 1.803 \\ 
124 & UGC08286 & 9.10 & 1.05 $\pm$ 0.07 & \dots & 6.45 $\pm$ 0.18 & 0.99 & 90.0 $\pm$ 1.8 & 1.00 & 2.637 \\ 
125 & UGC07399 & 9.06 & 0.59 $\pm$ 0.10 & \dots & 15.2 $\pm$ 1.2 & 1.81 & 58.1 $\pm$ 2.6 & 1.06 & 1.895 \\ 
126 & NGC4214 & 9.06 & 0.46 $\pm$ 0.11 & \dots & 2.87 $\pm$ 0.14 & 1.00 & 18.5 $\pm$ 1.2 & 1.23 & 1.062 \\ 
127 & UGC05414 & 9.05 & 0.41 $\pm$ 0.09 & \dots & 7.45 $\pm$ 0.82 & 0.79 & 54.6 $\pm$ 3.0 & 0.99 & 1.299 \\ 
128 & UGC08490 & 9.01 & 0.86 $\pm$ 0.11 & \dots & 4.81 $\pm$ 0.35 & 1.03 & 50.7 $\pm$ 2.4 & 1.01 & 0.337 \\ 
129 & IC2574 & 9.01 & 0.07 $\pm$ 0.00 & \dots & 3.78 $\pm$ 0.19 & 0.97 & 64.5 $\pm$ 3.4 & 0.86 & 1.440 \\ 
130 & UGC06446 & 8.99 & 1.04 $\pm$ 0.17 & \dots & 11.7 $\pm$ 1.2 & 0.97 & 50.9 $\pm$ 2.9 & 1.00 & 0.996 \\ 
131 & F583-1 & 8.99 & 0.91 $\pm$ 0.14 & \dots & 31.9 $\pm$ 3.5 & 0.90 & 62.3 $\pm$ 4.9 & 0.99 & 2.663 \\ 
132 & UGC11820 & 8.99 & 1.01 $\pm$ 0.11 & \dots & 15.5 $\pm$ 4.2 & 0.85 & 38.7 $\pm$ 6.7 & 0.86 & 1.988 \\ 
133 & UGC07690 & 8.93 & 0.60 $\pm$ 0.13 & \dots & 6.90 $\pm$ 1.44 & 0.85 & 39.4 $\pm$ 4.4 & 0.96 & 1.525 \\ 
134 & UGC04305 & 8.87 & 0.71 $\pm$ 0.16 & \dots & 3.44 $\pm$ 0.17 & 1.00 & 18.3 $\pm$ 0.9 & 0.46 & 2.024 \\ 
135 & NGC2915 & 8.81 & 0.32 $\pm$ 0.05 & \dots & 4.68 $\pm$ 0.17 & 1.15 & 70.3 $\pm$ 2.7 & 1.26 & 4.017 \\ 
136 & UGC05716 & 8.77 & 1.41 $\pm$ 0.07 & \dots & 18.6 $\pm$ 4.0 & 0.87 & 48.6 $\pm$ 7.2 & 0.90 & 5.664 \\ 
137 & UGC05829 & 8.75 & 0.60 $\pm$ 0.14 & \dots & 8.03 $\pm$ 2.28 & 0.93 & 29.4 $\pm$ 5.5 & 0.86 & 0.454 \\ 
138 & F565-V2 & 8.75 & 0.50 $\pm$ 0.12 & \dots & 56.7 $\pm$ 6.9 & 1.09 & 64.2 $\pm$ 7.2 & 1.07 & 0.474 \\ 
139 & DDO161 & 8.74 & 0.23 $\pm$ 0.04 & \dots & 4.84 $\pm$ 0.97 & 0.65 & 66.3 $\pm$ 9.8 & 0.95 & 1.468 \\ 
140 & DDO170 & 8.73 & 0.79 $\pm$ 0.15 & \dots & 9.48 $\pm$ 1.47 & 0.62 & 63.8 $\pm$ 7.3 & 0.97 & 4.917 \\ 
141 & NGC1705 & 8.73 & 1.22 $\pm$ 0.13 & \dots & 6.23 $\pm$ 0.26 & 1.09 & 86.9 $\pm$ 3.8 & 1.09 & 0.373 \\ 
142 & UGC05721 & 8.73 & 0.62 $\pm$ 0.08 & \dots & 8.66 $\pm$ 0.75 & 1.40 & 63.7 $\pm$ 4.3 & 1.04 & 1.824 \\ 
143 & UGC08837 & 8.70 & 0.20 $\pm$ 0.03 & \dots & 6.39 $\pm$ 0.38 & 0.89 & 58.4 $\pm$ 4.2 & 0.73 & 2.349 \\ 
144 & UGC07603 & 8.58 & 0.34 $\pm$ 0.06 & \dots & 6.55 $\pm$ 0.42 & 1.39 & 78.4 $\pm$ 3.0 & 1.01 & 1.772 \\ 
145 & UGC00891 & 8.57 & 0.32 $\pm$ 0.07 & \dots & 9.24 $\pm$ 0.99 & 0.91 & 59.5 $\pm$ 4.8 & 0.99 & 25.160 \\ 
146 & UGC01281 & 8.55 & 0.39 $\pm$ 0.06 & \dots & 5.27 $\pm$ 0.20 & 1.00 & 90.0 $\pm$ 0.6 & 1.00 & 0.244 \\ 
147 & UGC09992 & 8.53 & 0.51 $\pm$ 0.13 & \dots & 9.62 $\pm$ 3.11 & 0.90 & 20.9 $\pm$ 5.1 & 0.70 & 1.076 \\ 
148 & D512-2 & 8.51 & 0.48 $\pm$ 0.12 & \dots & 12.7 $\pm$ 3.1 & 0.83 & 51.7 $\pm$ 8.5 & 0.92 & 0.370 \\ 
149 & UGC00731 & 8.51 & 2.39 $\pm$ 0.45 & \dots & 9.45 $\pm$ 0.79 & 0.76 & 56.6 $\pm$ 3.0 & 0.99 & 6.415 \\ 
150 & UGC08550 & 8.46 & 0.74 $\pm$ 0.11 & \dots & 6.32 $\pm$ 0.32 & 0.94 & 90.0 $\pm$ 1.8 & 1.00 & 1.552 \\ 
151 & UGC07608 & 8.42 & 0.48 $\pm$ 0.12 & \dots & 8.57 $\pm$ 2.12 & 1.04 & 27.9 $\pm$ 4.5 & 1.12 & 0.734 \\ 
152 & NGC2366 & 8.37 & 0.24 $\pm$ 0.03 & \dots & 3.09 $\pm$ 0.16 & 0.95 & 54.9 $\pm$ 2.3 & 0.81 & 1.934 \\ 
153 & NGC4068 & 8.37 & 0.38 $\pm$ 0.09 & \dots & 4.30 $\pm$ 0.22 & 0.98 & 31.7 $\pm$ 1.9 & 0.72 & 2.519 \\ 
154 & UGC05918 & 8.37 & 0.54 $\pm$ 0.13 & \dots & 6.63 $\pm$ 1.22 & 0.87 & 44.8 $\pm$ 4.5 & 0.97 & 0.936 \\ 
155 & D631-7 & 8.29 & 0.20 $\pm$ 0.04 & \dots & 7.53 $\pm$ 0.18 & 0.98 & 45.9 $\pm$ 1.1 & 0.78 & 15.872 \\ 
156 & NGC3109 & 8.29 & 0.21 $\pm$ 0.04 & \dots & 1.43 $\pm$ 0.05 & 1.07 & 76.0 $\pm$ 3.5 & 1.09 & 4.133 \\ 
157 & UGCA281 & 8.29 & 0.37 $\pm$ 0.06 & \dots & 5.45 $\pm$ 0.26 & 0.96 & 64.5 $\pm$ 2.8 & 0.96 & 0.469 \\ 
158 & DDO168 & 8.28 & 0.46 $\pm$ 0.11 & \dots & 4.12 $\pm$ 0.21 & 0.97 & 50.3 $\pm$ 2.1 & 0.80 & 19.714 \\ 
159 & DDO064 & 8.20 & 0.48 $\pm$ 0.11 & \dots & 6.21 $\pm$ 0.83 & 0.91 & 59.6 $\pm$ 4.8 & 0.99 & 0.334 \\ 
160 & PGC51017 & 8.19 & 0.44 $\pm$ 0.10 & \dots & 3.19 $\pm$ 0.37 & 0.23 & 63.4 $\pm$ 3.3 & 0.96 & 4.567 \\ 
161 & UGCA442 & 8.15 & 0.44 $\pm$ 0.10 & \dots & 4.35 $\pm$ 0.20 & 1.00 & 64.1 $\pm$ 3.2 & 1.00 & 7.650 \\ 
162 & UGC07866 & 8.09 & 0.45 $\pm$ 0.11 & \dots & 4.48 $\pm$ 0.23 & 0.98 & 34.5 $\pm$ 2.4 & 0.79 & 0.260 \\ 
163 & UGC07232 & 8.05 & 0.46 $\pm$ 0.09 & \dots & 2.83 $\pm$ 0.16 & 1.00 & 59.1 $\pm$ 4.2 & 1.00 & 6.169 \\ 
164 & UGC07559 & 8.04 & 0.31 $\pm$ 0.06 & \dots & 4.43 $\pm$ 0.24 & 0.89 & 51.4 $\pm$ 2.6 & 0.84 & 2.602 \\ 
165 & NGC6789 & 8.00 & 0.60 $\pm$ 0.14 & \dots & 3.60 $\pm$ 0.17 & 1.02 & 53.9 $\pm$ 4.9 & 1.25 & 5.904 \\ 
166 & KK98-251 & 7.93 & 0.44 $\pm$ 0.10 & \dots & 3.35 $\pm$ 0.47 & 0.49 & 57.4 $\pm$ 5.2 & 0.97 & 1.227 \\ 
167 & UGC05764 & 7.93 & 3.83 $\pm$ 0.50 & \dots & 7.14 $\pm$ 1.32 & 0.96 & 59.3 $\pm$ 8.3 & 0.99 & 16.177 \\ 
168 & CamB & 7.88 & 0.34 $\pm$ 0.08 & \dots & 2.83 $\pm$ 0.30 & 0.84 & 26.9 $\pm$ 2.3 & 0.41 & 5.758 \\ 
169 & ESO444-G084 & 7.85 & 0.42 $\pm$ 0.09 & \dots & 5.08 $\pm$ 0.43 & 1.05 & 40.1 $\pm$ 2.2 & 1.25 & 3.253 \\ 
170 & DDO154 & 7.72 & 0.19 $\pm$ 0.03 & \dots & 3.87 $\pm$ 0.16 & 0.96 & 61.2 $\pm$ 2.1 & 0.96 & 3.482 \\ 
171 & UGC07577 & 7.65 & 0.24 $\pm$ 0.05 & \dots & 2.14 $\pm$ 0.14 & 0.83 & 45.5 $\pm$ 2.7 & 0.72 & 5.794 \\ 
172 & D564-8 & 7.52 & 0.40 $\pm$ 0.09 & \dots & 8.69 $\pm$ 0.28 & 0.99 & 42.5 $\pm$ 2.4 & 0.67 & 3.160 \\ 
173 & NGC3741 & 7.45 & 0.31 $\pm$ 0.05 & \dots & 3.35 $\pm$ 0.12 & 1.04 & 72.8 $\pm$ 3.1 & 1.04 & 0.767 \\ 
174 & UGC04483 & 7.11 & 0.43 $\pm$ 0.10 & \dots & 2.55 $\pm$ 0.22 & 0.76 & 53.0 $\pm$ 2.9 & 0.91 & 0.869 \\ 
175 & UGCA444 & 7.08 & 0.49 $\pm$ 0.12 & \dots & 0.84 $\pm$ 0.04 & 0.86 & 67.2 $\pm$ 4.0 & 0.86 & 0.330 \\ 
\end{longtable}
}
\clearpage
%\include{FigureSet}
%\end{appendix}

\end{CJK*} 

\end{document}